\documentclass[onecolumn,secnumarabic,amssymb,nobibnotes,aps,showpacs,nofootinbib,noeprint,superscriptaddress,floatfix]{revtex4-1}
\usepackage{graphicx} % Required for inserting images

\usepackage{amsmath}
\usepackage{amssymb}
\usepackage[utf8]{inputenc}
\usepackage{graphicx}
\usepackage[english]{babel}
\usepackage{setspace}
\usepackage{hyperref}
\usepackage{etoolbox}
\usepackage{xcolor}
\usepackage{subcaption}
\usepackage{ragged2e}
\usepackage{geometry}

\usepackage{lmodern}

\newcommand{\angstrom}{\text{\normalfont\AA}}

\usepackage{appendix}
\newcommand{\Sc}{\operatorname{\mathit{S\kern-.07em c}}}
\newcommand{\Pe}{\operatorname{\mathit{P\kern-.08em e}}}

\patchcmd{\thebibliography}{\section*{\refname}}{}{}{}
\begin{document}
	%\sffamily
    \rmfamily
	\begin{spacing}{1.25}

\title{Flow-induced anisotropy in a carbon black-filled silicone elastomer:
electromechanical properties and structure}
\author{Bettina Zimmer}
\affiliation{INM – Leibniz Institute for New Materials, Campus D2 2, 66123 Saarbrücken, Germany}
\author{Bart-Jan Niebuur}
\affiliation{INM – Leibniz Institute for New Materials, Campus D2 2, 66123 Saarbrücken, Germany}
\author{Florian Schaefer}
\affiliation{Materials Science and Methods, Saarland University, Campus D2 3/C5 3, 66123 Saarbruecken, Germany}
\author{Fabian Coupette}
\affiliation{Institute of Physics, University of Freiburg, Hermann-Herder-Stra{\ss}e 3, 79104 Freiburg, Germany}
\affiliation{School of Mathematics, University of Leeds, Leeds LS2 9JT, United Kingdom}
\author{Victor T\"anzel} 
\affiliation{Institute of Physics, University of Freiburg, Hermann-Herder-Stra{\ss}e 3, 79104 Freiburg, Germany}
\affiliation{Cluster of Excellence \textit{liv}MatS @ FIT – Freiburg Center for Interactive Materials and Bioinspired Technologies, University of Freiburg, Georges-Köhler-Allee 105, D-79110 Freiburg, Germany}
\author{Tanja Schilling}
\affiliation{Institute of Physics, University of Freiburg, Hermann-Herder-Stra{\ss}e 3, 79104 Freiburg, Germany}
\author{Tobias Kraus}
\affiliation{INM – Leibniz Institute for New Materials, Campus D2 2, 66123 Saarbrücken, Germany}
\affiliation{Colloid and Interface Chemistry, Saarland University, Campus D2 2, 66123 Saarbrücken, Germany}
\date{\today}

\begin{abstract}

\noindent \textbf{Abstract}
 
\noindent Carbon black (CB)-elastomers can serve as low-cost, highly deformable sensor materials, but hardly any work exists on their structure-property relationships. We report on flow-induced anisotropy, considering CB-silicone films generated via doctor blade coating.
Cured films showed slight electrical anisotropy, with conductivity parallel to the coating direction being lower than perpendicular to it. Furthermore, piezoresistive sensitivity was much larger for stretch perpendicular to the coating direction than for parallel stretch.
Structural analysis for length scales up to the CB agglomerate level yielded only weak evidence of anisotropy. Based on this evidence and insight from CB network simulations, we hypothesize that shear flow during coating fragments the CB network and then induces a preferential aggregate  alignment, as well as increased inter-particle distances, parallel to the coating direction. As a practical conclusion, already weak anisotropic structuration suffices to cause significant electric anisotropy.

\end{abstract}

\maketitle
\section{Introduction} \label{introduction}

Carbon black (CB) is a common filler used to tune mechanical properties of rubbers \cite{Araby-Elastomeric,Flandin-Effect,Karuthedath-Characterization,Kost-Effects,Li-A-review}. In addition, it introduces electrical conductivity when its concentration exceeds a critical value (percolation threshold) needed to form a network within the polymeric matrix \cite{Balberg-Anisotropic,Huang-Carbon,Kost-Resistivity,Okano-Anisotropic}. CB-filled elastomers can thus be used to create mechanically robust electronic devices such as flexible electrodes and highly deformable piezoresistive sensors \cite{Karuthedath-Characterization,Knite-Polyisoprene,Knite-Reversible,Panahi-Sarmad-A-comprehensive,Wang-Highly,Zavickis-Polyisoprene-nanostructured}. Compared to conductive fillers like carbon nanotubes (CNTs) \cite{Cattin-Piezoresistance,Chen-Acid,Nankali-Highly}, graphene \cite{Chen-PDMSbased,Shi-Highly}, and silver nanoparticles \cite{Feng-Highly,Lee-A-stretchable}, CB is an attractive alternative in that it is much less expensive and more readily available \cite{Araby-Elastomeric}.

To optimize CB elastomers for electronic applications, the correlation between structure (esp. CB dispersion state) and electromechanical properties must be understood. Literature on CB composites reports on numerous relevant factors, e.g. CB primary structure and surface chemistry \cite{Park-Influence-of-surface, Datta-Surface, Spahr-CarbonBlack, Kost-Effects}, CB aggregate size distribution \cite{Coupette-Percolation,Flandin-Anomalous}, CB-matrix interactions \cite{Miyasaka-Electrical}, additives (e.g. ionic liquids \cite{Zhang-Microscopic,Sattar-Surface}, inorganic salts \cite{Flandin-Anomalous}, non-ionic plasticizer \cite{Knite-The-effect}), matrix viscosity during processing \cite{Rwei-Dispersion}, and curing temperature \cite{Rwei-Dispersion}. In addition, the processing method has a decisive impact \cite{Araby-Elastomeric,Flandin-Effect,Karuthedath-Characterization,Balberg-Anisotropic,Huang-Carbon,Ehrburger-Dolle-Role,Huang-Strain,Leboeuf-Influence,Vigueras-Santiago-Electric,Zhang-Anisotropically}. Depending on the flow history of the precursor (mixture of CB and the yet liquid matrix), different CB morphologies can form, incl. anisotropic ones. The latter result from material flow in preferential directions, e.g. during injection molding \cite{Vigueras-Santiago-Electric}, compression molding \cite{Flandin-Effect,Balberg-Anisotropic}, melt-casting \cite{Ehrburger-Dolle-Anisotropic}, and extrusion \cite{Zhang-Anisotropically}. 

While extensive research on flow-induced anisotropy has been done on CB suspended in low viscosity organic liquids \cite{Grenard-Shear,Negi-New,Osuji-Shear,Osuji-Highly} and CB-filled thermoplasts \cite{Balberg-Anisotropic,Vigueras-Santiago-Electric,Zhang-Anisotropically,Ehrburger-Dolle-Anisotropic}, literature on the microstructure-dependent electromechanical properties of CB elastomers is scarce. We did not find any work for CB elastomers with chemically crosslinked matrices (e.g. silicone rubber) and only two publications \cite{Flandin-Effect,Ehrburger-Dolle-Anisotropic} for CB-filled thermoplastic elastomers (thermoplastic matrix with elastomeric properties). Flandin et al. \cite{Flandin-Effect} briefly discussed the possible role of compression molding in the electrical anisotropy of a CB-filled ethylene-octene elastomer in the undeformed state, but due to the absence of structural analysis, the structure-property relationship was not clarified. Ehrburger-Dolle et al. \cite{Ehrburger-Dolle-Anisotropic} reported anisotropic X-ray scattering patterns of ethylene propylene rubber with CB contents slightly above the percolation threshold. The anisotropy, which presumably originates from the liquid composite being sheared during melt-casting, is correlated with different degrees of interpenetration of CB aggregates in two principal orientations. Yet, no link between the principal orientations in the patterns and the flow conditions is established, and the consequences for electrical conductivity are not addressed.

In light of this lack, this work explores flow-induced anisotropy of a CB-filled silicone elastomer in terms of electromechanical properties and structure. For this, we exposed the yet liquid CB-silicone mixtures to shear flow by means of doctor blade coating at two gap heights (60 µm, 350 µm) and various coating speeds. The resulting films were cured at elevated temperature to retain the process-induced microstructure. Electrical resistance of cured, unstrained films of three CB concentrations above the percolation threshold was measured both parallel and perpendicular to the coating direction, and the influence of process parameters (blade speed, film thickness) was investigated (Section \ref{Local electrical properties (unstrained state)}). In addition, piezoresistivity (electrical resistance change under mechanical deformation) was studied via uniaxial tensile tests with in-situ electrical resistance measurement along the stretch axis (parallel or perpendicular to the coating direction, Section \ref{Electromechanical properties_results}). 

Shearing the uncured CB-silicone mixtures lead to significant electrical and piezoresistive anisotropy in the cured material, whereas mechanical anisotropy was negligible. We discuss implications of this (piezo-)electric anisotropy for industrial practice and explore its structural origin via characterization of unstretched and stretched states by small-angle X-ray scattering (SAXS) as well as nanomechanical mapping (PeakForce QNM) combined with segmentation of the signal maps (Section \ref{Structural analysis}). In addition, we present simulations of the fractal filler network, namely on the impact  of shear flow on particle alignment and consequences for electrical anisotropy (Section \ref{Simulations_results}). A structural hypothesis for the observed phenomenology is given (Section \ref{structural hypothesis}) which, due to the rather indirect nature of the structural evidence, needs to be validated in future work.

\section{Experimental} \label{Experimental}
\subsection{Film fabrication and sample preparation} \label{Fabrication of films}

CB-filled elastomeric films of at least 5x7 cm\textsuperscript{2} were fabricated in ambient air by dispersing CB (7/9/11 vol\% CB, above the percolation threshold of $\sim$5 vol\% \cite{Coupette-Percolation}) in a crosslinking silicone matrix (Sylgard\textsuperscript{®} 184, Dow\textsuperscript{®}), doctor blade coating the resulting reactive mixtures (two gap heights, $h_{gap}$ = 60 µm and 350 µm), and curing for 2~h at 100~°C. In order to assess the impact of dynamic effects during processing on final properties, the blade speed, $v_{blade}$, was varied between 5~mm/s and 400~mm/s. The resulting global shear rates ($\gamma_{global}$= $v_{blade}$/$h_{gap}$) span a large interval of 50~s$^{-1}$ to 3000~s$^{-1}$. More details on film fabrication are found in Appendix \ref{Film fabrication (detailed description)}.

Cured, unstrained films were characterized electrically by the four-point method (Section \ref{Electrical resistance: four-point probe measurement}) without any further sample preparation. Samples characterized by further methods (tensile test with in situ electrical two-point measurement, PeakForce QNM and SAXS, films coated at $h_{gap}$ = 350~µm, 20~mm/s) were cut from near the edges of the films as exemplified in Fig. \ref{fig:tensile & 4PP specimens, 4PP polarity, planes+directions}a (left/right side for stretch parallel to the coating direction, start/end position for stretch perpendicular to the coating direction). Exact sample dimensions are given in the corresponding sections (\ref{Electromechanical properties_experimental}, \ref{PeakForce QNM (quantitative nanomechanics)_experimental} and \ref{SAXS_experimental}).

 \begin{figure}
     \centering
     \includegraphics[width=1\linewidth]{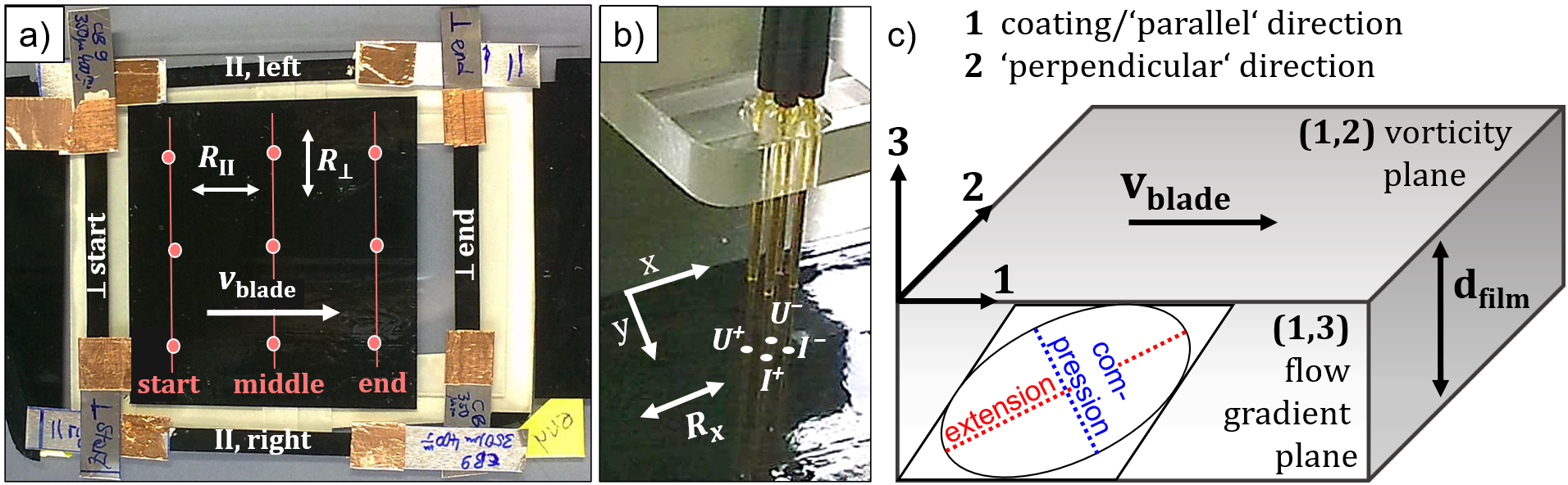}
     \caption{\justifying{To study anisotropy of CB-filled silicone films generated by doctor blade coating, samples were prepared and characterized parallel and perpendicular to the coating direction. a) film with tensile specimens cut from near the edges and example of measuring grid (spacing 3~–~4~cm) for electrical four-point measurements indicating ‘start’, ‘middle’, and ‘end’ positions, b) probe arrangement and polarity of four-point measurements in the square setup, c) definition of directions and planes with respect to the coating direction.}}
     \label{fig:tensile & 4PP specimens, 4PP polarity, planes+directions}
 \end{figure}

\subsection{Characterization} \label{Characterization}
\subsubsection{Four-point probe measurement} \label{Electrical resistance: four-point probe measurement}
Electrical resistance of undeformed films was measured in ambient air (22°C ± 1~K, 26~±~19~\%~r.h.) with a Keithley instrument (2450 Interactive SourceMeter®, current sweep from -10~µA to +10~µA, R obtained from linear fits of the ohmic voltage-current curves) using a customized four-point probe setup in the square arrangement (Fig. \ref{fig:tensile & 4PP specimens, 4PP polarity, planes+directions}b). With the used probe distance of 2 mm, all films can be approximated as thin films, giving the following analytical expressions for resistance, $R$, in two orthogonal directions $x$ and $y$ \cite{Miccoli-four}:

\begin{equation} \label{eq:Rx and Ry}
R_x=\frac{\sqrt{\rho_x\rho_y}}{2\pi\cdot d_{\mathrm{film}}}\cdot \ln(1+\rho_x/\rho_y) \\
\end{equation}
\begin{equation}
R_y=\frac{\sqrt{\rho_y\rho_x}}{2\pi\cdot d_{\mathrm{film}}}\cdot \ln(1+\rho_y/\rho_x)
\end{equation}

with $\rho_x, \rho_y$ the intrinsic resistivity in the x- and y-direction, respectively, and $d_{\mathrm{film}}$ the film thickness.

To measure resistances parallel and perpendicular to the coating direction, \(R_\parallel\) and \(R_\perp\), the films were arranged such that the latter was parallel or perpendicular to \(R_x\), respectively (see Fig. \ref{fig:tensile & 4PP specimens, 4PP polarity, planes+directions}b). When oriented in parallel, $R_x$ equals \(R_\parallel\); when oriented perpendicular, it equals \(R_\perp\).

Spatial variations were analyzed by choosing three to four measuring spots at the ‘start’, ‘middle’ and ‘end’ sections of each film, as schematically depicted in Fig. \ref{fig:tensile & 4PP specimens, 4PP polarity, planes+directions}a. The respective mean and maximal error of \(R_\parallel\) and \(R_\perp\) were then calculated from the three to four spots per section (one measurement per spot for each orientation). Note that the spots at a given grid position were not exactly the same for the parallel and perpendicular direction since the measuring direction was switched by rotating the film (rather than the probe polarity).

According to eq. \eqref{eq:Rx and Ry}, the resistivities, $\rho_x$ and $\rho_y$, cannot be determined separately from measuring $R_x$ and $R_y$. Despite this analytical limitation, the quotient $R_\parallel$/$R_\perp$ is a suitable measure for material anisotropy since it grows monotonically with the ratio of the intrinsic resistivities, $\rho_x/\rho_y$ \cite{Miccoli-four}:
\begin{equation}
\frac{R_x}{R_y} =\frac{\ln(1+\rho_x/\rho_y)}{\ln(1+\rho_y/\rho_x)}	
\label{eq:ratio}
\end{equation}
\(R_x/R_y\), or in the context of the films, \(R_\parallel/R_\perp\), is referred to as ‘electrical anisotropy ratio’.

\subsubsection{Uniaxial tensile test with electrical two-point probe measurement} \label{Electromechanical properties_experimental}
To study piezoresistivity, uniaxial tensile tests (universal testing machine Zwick 1446, sample stiffness negligible compared to stiffness of the machine) with in-situ electrical two-point measurements (DAQ6510 by Keithley, constant test current in ‘auto range’ mode, = 10 µA for the measured samples) were performed in ambient air (22°C ± 1~K, 30~±~15~\% r.h.). As samples, rectangular strips of 4x55 mm\textsuperscript{2} were cut from ‘thick’ films ($h_{gap}$ = 350 µm) coated at 20 mm/s, both for stretch in the parallel (engineering strain $\epsilon = \epsilon_\parallel$) and the perpendicular direction ($\epsilon = \epsilon_\perp$), respectively (Fig. \ref{fig:tensile & 4PP specimens, 4PP polarity, planes+directions}a). For in-situ resistance measurement along the respective stretch axis, \(R_\parallel(\epsilon_\parallel)\) and \(R_\perp(\epsilon_\perp)\), aluminum strips were glued to the top and bottom of the samples with conductive silver glue (Elektrodag 1415 M by Plano) and fixated with copper foil. The resulting probe distance is identical to the gauge length for straining and equals $L_0$ = 35 mm in the unstretched state. Samples mounted in the testing machine were contacted electrically with crocodile clamps.

After mounting, the tensile force was zeroed and the sample stretched to a pre-load of 0.05~N. The tensile test started after both force and electrical resistance had stabilized (roughly after 20~–~25 min), with the force zeroed at the start of the loading phase. The testing procedure consisted of 4 load-unload cycles between 0 \% strain and maximal strains of $\epsilon_{max}$ = 10/20/30/40 \%, with a strain rate of 10$^{-2}$ s$^{-1}$ (controlled via the position of the crosshead). After each loading/unloading, the material was allowed to relax for 20 min at the given strain plateau. From these relaxation phases, relaxed resistance values for each strain plateau were derived as explained in Appendix \ref{appendix: electromechanical}.

To differentiate between the impact of doctor blade coating on the silicone matrix vs. on the CB network, neat Sylgard (0~vol\% CB, same blade speed, slightly higher gap height of 400 µm resulting in a similar film thickness, see Appendix \ref{Film fabrication (detailed description)}) was characterized in addition to the CB composites (7/9/11 vol\% CB). Mechanical testing was identical to the CB-filled samples except for a lower pre-load (0.02 N) since the unfilled material is much less stiff.

Resistance measured along the stretch axis, $R_{x}$, is related to intrinsic electrical resistivity, $\rho_{x}$, and the geometrical contribution (sample length, $L$, and cross-sectional area, $A$) according to the well-known equation
\begin{equation} \label{eq:resistance as function of stretch}
R_x(\epsilon_x)=\rho_x (\epsilon_x)L(\epsilon_x)/A(\epsilon_x)
\end{equation}

where the x-orientation is either parallel or perpendicular to the coating direction in our tests. Thus, in addition to intrinsic resistivity, the geometrical part can lead to piezoresistive anisotropy: Mechanical anisotropy in the form of direction-dependent transverse contraction (compressibility) leads to different cross-sectional areas for a given stretch.

\subsubsection{PeakForce QNM (quantitative nanomechanics)} \label{PeakForce QNM (quantitative nanomechanics)_experimental}
PeakForce QNM™ by Bruker is a mode of scanning force microscopy (SFM) for quantitative nanomechanical mapping. It outputs sample topography (‘height’ signal) and, thanks to real-time analysis of force-distance curves for each pixel on the sample surface, local material properties (mirrored by the signals ‘deformation’, ‘dissipation’, ‘modulus’ and ‘adhesion’). For more information on PeakForce QNM, the reader is referred to \cite{Hua-PeakForce-QNM,Pittenger-Nanoscale,Pittenger-Quantitative} and Appendix \ref{Appendix: PeakForce QNM maps and value distributions}.

Measurements were performed with Bruker’s Dimension Icon in ambient air (22.5°C ± 0.5~K, 34~±~5 \% r.h.). The most important methodological details are compiled in Table \ref{tab:PeakForce QNM measuring parameters}. Calibration for quantitative measurements (deflection sensitivity, sync distance, PFT amplitude sensitivity) was done on sapphire via the ‘touch calibration’ feature of the software (NanoScope 9.30). 

\begin{figure}
    \centering
    \includegraphics[width=1\linewidth]{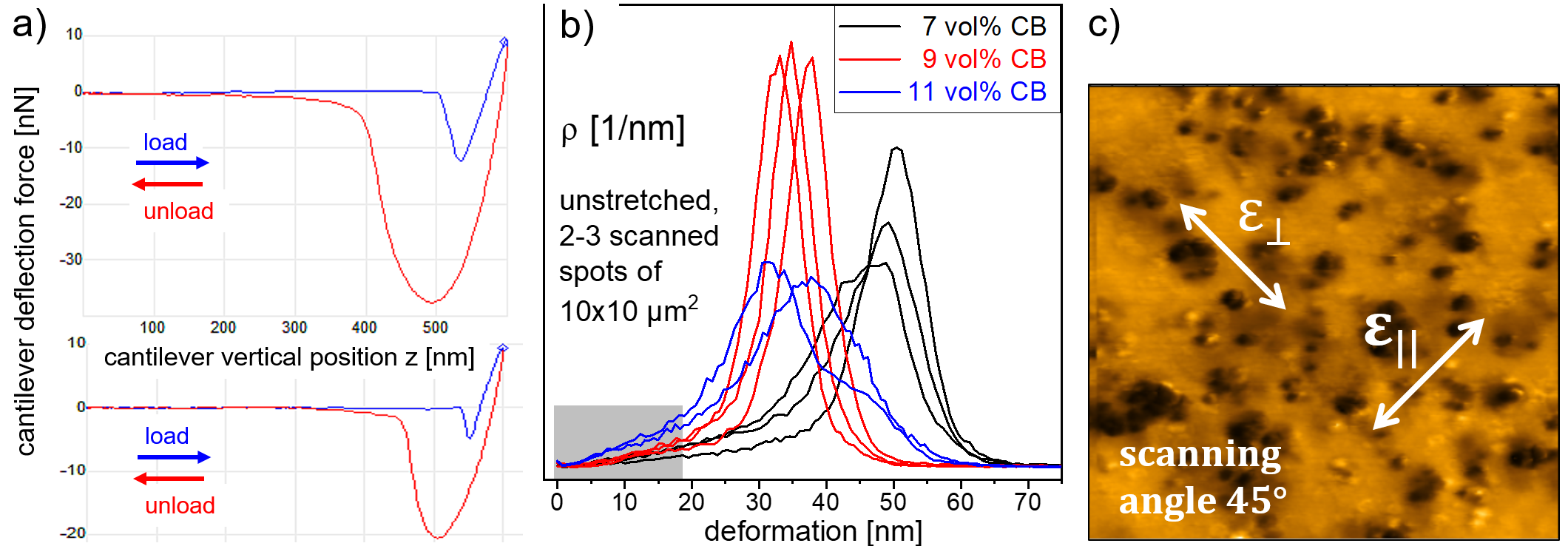}
    \caption{\justifying{PeakForce QNM revealed the morphology of near-surface CB. Suitable probe and measuring parameters ensured clean and reproducible data: a) Examples of force-distance curves (adapted from screenshots of real-time display in NanoScope),
b) frequency distributions of deformation signal (exported from Gwyddion 2.62) of unstretched films with 7/9/11 vol\% CB (grey shaded area: range dominated by CB particles, rest: range dominated by the silicone matrix), c) visualization of the 45° scanning angle which minimizes bias from the z-control of the cantilever (smearing of features in to the scanning direction) parallel and perpendicular to the coating direction.}}
    \label{fig:force-distance curves}
\end{figure}

For characterization in unstrained and strained (40 \% parallel/perpendicular to the coating direction) states, rectangular strips were cut from the 350~µm-films coated at 20 mm/s as shown in Section \ref{Fabrication of films}. Sample width ($\sim$1~cm) and length ($\sim$5~cm) were chosen big enough to ensure a center region suitable for scanning (no edge effects) and to allow fixation in the stretched state. Glass slides glued to the SFM stage using white-out served as a substrate. In the unstrained state, sample adhesion to the glass was sufficient for stable scanning. For the strained state, the strips were manually stretched to 40 \% strain, fixated with tape, and allowed to relax before characterization (at least 30 min). Samples were scanned on their bottom surface, i.e., the surface generated by the contact to the substrate foil during coating. In contrast to the top surfaces generated by the doctor blade, the bottom surfaces of all investigated compositions (7/9/11 vol\% CB) have a topography which allows high quality measurements (no serious artifacts from topological features; rms-roughness = 2~–~5~nm when unstretched). To test for drift during scanning, measuring spots were scanned at least twice. All data presented and discussed here is devoid of artifacts from sample drift. To minimize bias from the control of the vertical position of the cantilever (z), the scanning angle was set to 45° with respect to the coating direction. The parallel and perpendicular stretch axes in the resulting images are indicated in Fig. \ref{fig:force-distance curves}c. Any artifacts from the z-control during scanning (shape distortion of carbon black in particular) are thus equal for the parallel and the perpendicular orientation.

\begin{table}
    \centering
        \caption{Most important methodological information on the PeakForce QNM measurements}
    \label{tab:PeakForce QNM measuring parameters}
    \begin{tabular}{|c|c|}
          \hline
         SFM probe: name, material, & RTESPA-300-15 (pre-calibrated), Sb-doped Si,\\
         cantilever stiffness k, tip end radius R 
          & k = 3.379 N/m, R = 30 nm \\
         \hline
          scan size &	5x5/10x10/20x20 µm$^2$ \\
         \hline
         aspect ratio & 1 \\
         \hline
         scan angle & 45° \\
         \hline
         scan rate & 0.854 Hz for most scans \\
         & (other used values: 0.450 Hz or 0.976 Hz) \\
        \hline
         samples per line (pixel density) & 256 \\
         \hline
         PeakForce & 10 nN \\
         \hline
         oscillation frequency & 1 kHz \\
        \hline
         oscillation amplitude & 300 nm \\
        \hline
         minimum to maximum force fit boundary & 30 – 90 \% \\
         \hline
         deformation force level & 15 \% \\
        \hline
         software for data analysis and representation & Gwyddion 2.62 \\
          \hline
    \end{tabular}
\end{table}

Probe and measuring parameters were chosen to give clean force-distance curves on the soft areas dominated by the silicone matrix as well as on the much stiffer CB-rich regions (see Fig. \ref{fig:force-distance curves}a for examples). This involves sufficient indentation on stiff regions (for high quality force-distance curves) and minimal indentation on soft regions (corresponding to maximal lateral resolution). According to the frequency density of the deformation signal (see Fig. \ref{fig:force-distance curves}b as well as Appendix \ref{Appendix: PeakForce QNM maps and value distributions} for further explanation), the indentation depth varied between 0 nm and 70 nm for all examined CB contents (7/9/11 vol\%). Together with the tip end radius of 30 nm, this corresponds to contact radius values in the same range (a few 10\textsuperscript{1} nm). The chosen pixel density (256 per line) resulted in a pixel size suitable for these values (20/40/80~nm for scans of 5x5/10x10/20x20~µm\textsuperscript{2}). Note that the sample volume contributing to the material response exceeds the contact region and indentation depth. As a rule of thumb, it extends to some multiples of the contact radius (downward from the sample surface + laterally from the rotational axis of the tip, see e.g. \cite{Maugis-Contact} for calculations), which in our case equates to a probed depth of some 10\textsuperscript{1} – 10\textsuperscript{2} nm. As indicated by the shaded area in Fig. \ref{fig:force-distance curves}b (see Appendix \ref{Appendix: PeakForce QNM maps and value distributions} for discussion), regions dominated by CB are much less deformable than the ones dominated by the silicone matrix, with indentation depths of a few nm to about 20 nm. As evidenced by our results presented in Section \ref{Nanomechanical mapping (PeakForce QNM)_results}, the resulting lateral resolution suffices to discriminate CB aggregates.
     
\subsubsection{Segmentation and statistical analysis of PeakForce QNM data} \label{Segmentation and statistical analysis of PeakForce QNM data_experimental}

A particle analysis was carried out with ImageJ and Fiji 2.15.8, respectively \cite{schneider2012nih,schindelin2012fiji}. For this purpose, the dissipation signals of the PeakForce QNM measurements were first converted into 8-bit grayscale images using Gwyddion (2.64, Delayed Drifter).

The images were then segmented using numerical gray value thresholding. The threshold was set using the \textit{MaxEntropy} algorithm, implemented in Fiji. The filter uses the entropy of the gray value histogram derived on the basis of information theory to determine a threshold value \cite{kapur1985new}. The basic challenge is that the images show both details of particles that are exposed on the surface (black, also visible in the adhesion image) and of particles that are covered by matrix elastomer (dark to light gray). 

\begin{figure}
     \centering
     \includegraphics[trim={2cm 12cm 2cm 1cm},clip,width=0.95\linewidth]{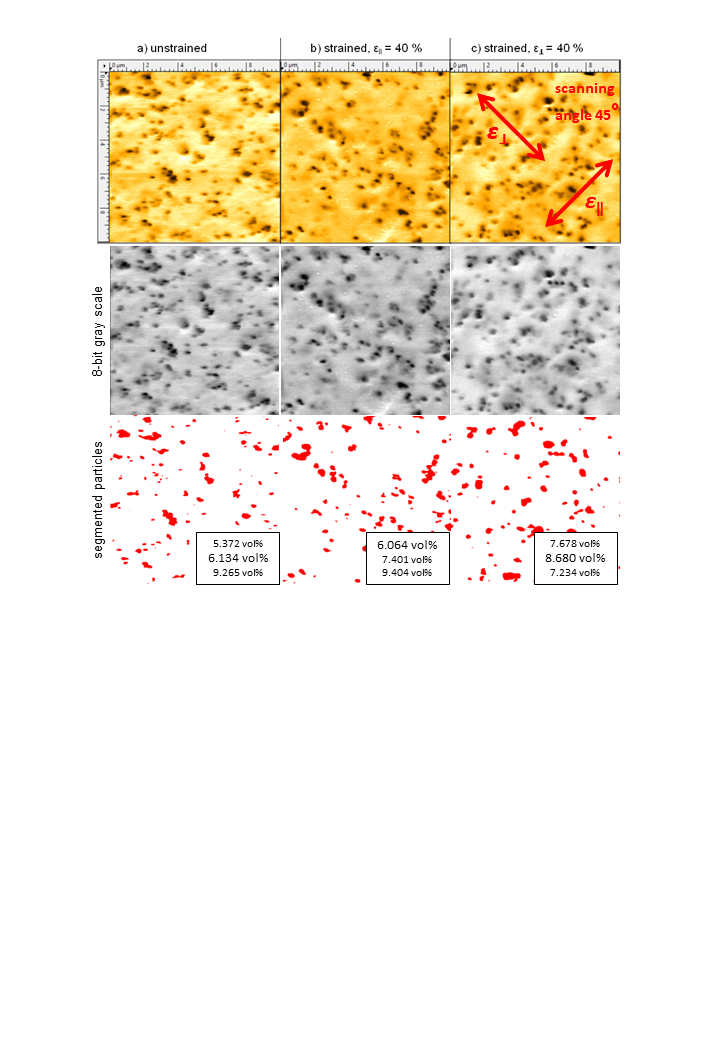}
     \caption{\justifying{Dissipation data from PeakForce QNM measurements for specimens with 7 vol\% CB in the unstretched state and at 40~\% strain (parallel and perpendicular to the coating direction) are converted to 8-bit grayscale images and segmented by gray value thresholding using the \textit{MaxEntropy} algorithm implemented in Fiji. The images below illustrate the masks for particle detection. The inset shows the determined area and thus volume fractions of the particles, whereby 3 data sets were evaluated in each case. The enlarged value is the one that belongs to the data set shown.}}
     \label{fig:threshold}
 \end{figure}

The volume fraction $V_V$ of the CB particles is evaluated using the approach $V_V=A_A$, where $A_A$ denotes the fraction of particles in a perfect two-dimensional surface section through the bulk sample \cite{ohser2000statistical}, assuming a uniform density of CB particles in the film. %The critical point here is that many particles are pressed into the matrix during coating or are covered with elastomer coincident with possible sedimentation. The latter might lead to an overestimation of the particles' volume fraction by image analysis.  This is counteracted by the fact that PeakForce QNM dissipation signals have an information depth in the order of magnitude of the agglomerates, so that particles below the surface also contribute to the surface signal, though with brighter gray values with increasing depth.  Nevertheless, 
In overview scans to 20x20 ~$\mu m^2$, the threshold determination with the \textit{MaxEntropy} filter proved successful, as it gave a good approximation of the volume fraction of samples with 7/9/11~vol\% CB, with a tendency to slight overestimation. 
The volume fractions of the CB particles measured in 10x10 µm\textsuperscript{2} images of the 7~vol\% film are indicated in Fig. \ref{fig:threshold}. 

This is followed by erosion and dilation by one pixel each to achieve a better discrimination of CB aggregates, as SFM scans tend to smear the edges of raised features in the scan direction due to control delay. The volume fraction is then determined again. This is followed by the actual particle analysis: First, the angle for each particle in which the maximum Feret diameter lies with respect to the horizontal direction is determined and this is plotted as a histogram over all particles from 3 scans each for the same parameter. (The coating direction is at 45$^\circ$ with respect to the horizontal direction). The average particle diameter $2\sqrt{\mathrm{area}/\pi}$, the particle size distribution and the average particle distance as well as the circularity of the particles as $\frac{4 \pi \mathrm{area}}{\mathrm{perimeter}^2}$ are determined as well, see Fig. \ref{fig:segmentation results_edit BZ}. The circularity approaches 1.0 for perfectly round particles and 0.0 for elongated particles. Particles that are cut by the edges of the image are excluded from the evaluation. 

\subsubsection{SAXS} \label{SAXS_experimental}
SAXS measurements were performed on a laboratory-scale Xeuss 2.0 instrument (Xenocs SA, Grenoble, France). The X-ray beam from a copper K$_{\alpha}$ source (wavelength $1.54\,\angstrom$) was focused on the sample with a spot size of $0.25\,\text{mm}^{2}$. The samples were located at a sample-detector distance (SDD) of $2500~\text{mm}$, calibrated using a silver behenate standard. The resulting measurable momentum transfer, $q$, ranges from $5\cdot10^{-3}\angstrom^{-1}$ to $2\cdot10^{-1}\angstrom^{-1}$, with $q$ being defined as $q = 4\pi \text{sin}(\theta/2)/\lambda$ and $\theta$ the scattering angle. 2D scattering patterns were obtained using a Pilatus 300K detector (Dectris, Baden, Switzerland) with a pixel size of 0.172 × 0.172 mm$^{2}$ and an acquisition time of $1\,\text{h}$ for each sample.

The samples (0~vol\% CB coated at 400~µm, 20~mm/s and 9~vol\% CB coated at 350 µm, ~20~mm/s, $\sim$7x50 mm\textsuperscript{2}-strips cut from near the film edges parallel and perpendicular to the coating direction, respectively) were placed directly in the beam, without the need of using a sample container. Sample strain is induced by manually stretching and fixing the samples at 140 \% of their original length (nominal strain of 40 \%).

To obtain $I$($q$) parallel and perpendicular to the direction of strain, the 2D scattering patterns were azimuthally averaged within two angle ranges subtending 20 $\deg$ parallel and perpendicular to direction of strain, respectively.

\section{Results and Discussion} \label{Results and Discussion}
\subsection{Electrical anisotropy of unstrained films} \label{Local electrical properties (unstrained state)}
\begin{figure}
     \centering
     \includegraphics[width=1\linewidth]{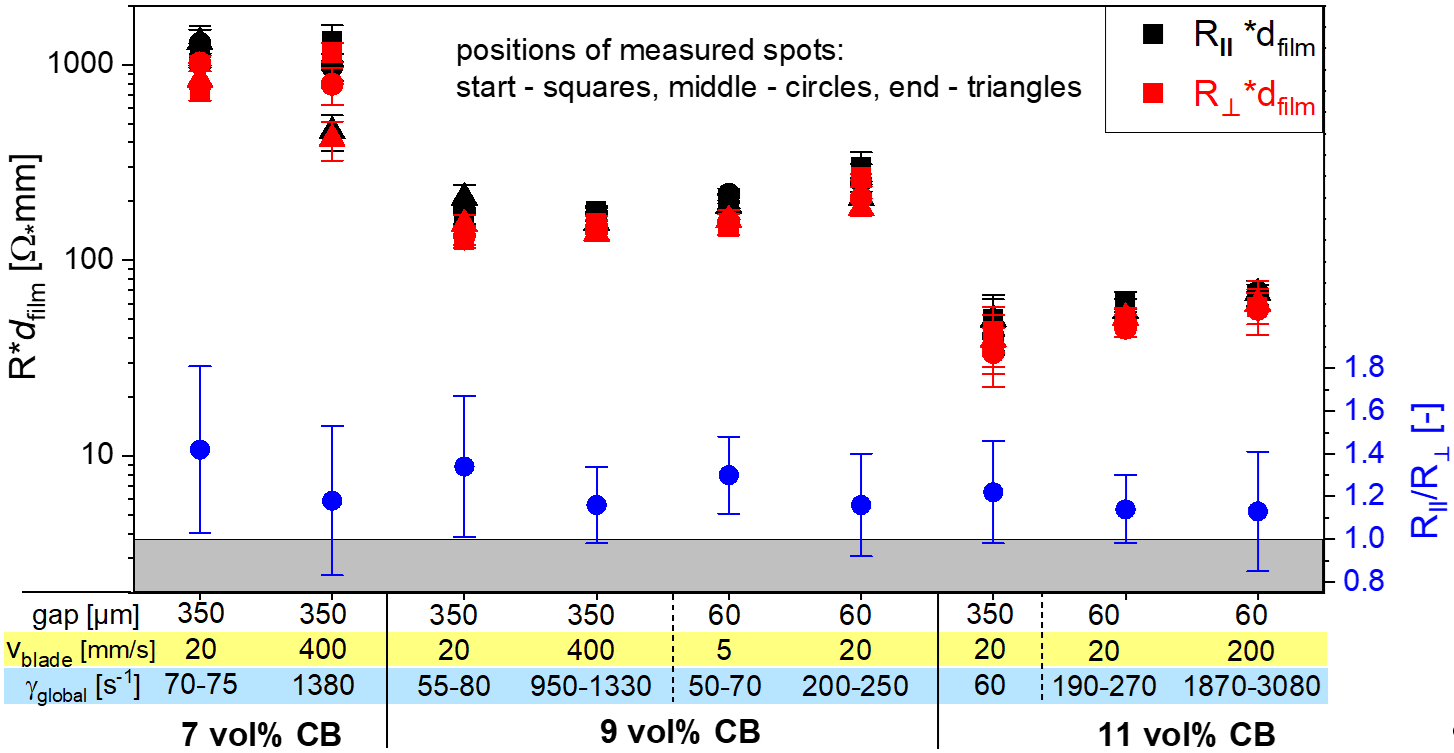}
     \caption{\justifying{Electrical four-point probe measurements (22°C ± 1 K, 26 ± 19 \% r.h.) reveal that cured, unstrained CB-silicone films produced by doctor blade coating (7/9/11 vol\% CB, gaps 60/350~µm, blade speeds 5~–~400~mm/s) are slightly less conductive in the coating direction that perpendicular to it. Shown in the figure are the respective resistances parallel and perpendicular to the coating direction, $R_\parallel$ and $R_\perp$ (multiplied by film thickness to compensate geometric variations, see eq. \eqref{eq:Rx and Ry} in Section \ref{Electrical resistance: four-point probe measurement}), measured at the start, middle and end position of the films (mean and maximal error from 3~–~4 points on each position), and the resulting anisotropy ratio, $R_\parallel$/$R_\perp$. The mean and maximal error of $R_\parallel$/$R_\perp$ were calculated from all local pairs of $R_\parallel$ and $R_\perp$ of a given film (9~–~12 measuring spots, see Section \ref{Electrical resistance: four-point probe measurement} for visualization of the measuring grid).}}
     \label{fig:Electrical resistances & anisotropy ratio_unstretched}
 \end{figure}

Electrical resistances of unstrained films in the two main orientations relative to the coating direction, $R_\parallel$ and $R_\perp$, along with corresponding values of the electrical anisotropy ratio, $R_\parallel$/$R_\perp$, are illustrated in Fig. \ref{fig:Electrical resistances & anisotropy ratio_unstretched}. Resistance values are multiplied by the corresponding film thickness to eliminate geometrical effects and facilitate comparison (see eq. \eqref{eq:Rx and Ry} in Section \ref{Electrical resistance: four-point probe measurement}). Symbols indicate the start, middle and end sections of film coatings. We found no relevant variation of resistances along the coating direction with one exception (7 vol\% CB, 400 mm/s). Doctor blade coating results in electrically uniform samples for the other parameters.

Resistance values parallel to the coating direction were systematically higher than perpendicular to it, resulting in $R_\parallel$/$R_\perp>1$ (mean: 1.1 – 1.4) for all films. Doctor blade coating the uncured CB-silicone mixtures apparently introduced electrical anisotropy, decreasing conductivity in the coating direction relative to the perpendicular direction. Within experimental error, the degree of electrical anisotropy depended neither on CB concentration nor on film thickness or blade speed. This is surprising in that all these parameters have a big impact on hydrodynamics: Increasing CB content from 7 vol\% to 11 vol\% leads to much more viscous mixtures, as evidenced by handling them during the coating process, and the variations of film thickness and blade speed imply a large range of global shear rates of approximately 50 – 3080 s\textsuperscript{-1} (see Fig. \ref{fig:Electrical resistances & anisotropy ratio_unstretched}).

Flow-induced electrical anisotropy in carbon filler-polymer composites has been reported previously. For example, CB-filled thermoplasts \cite{Vigueras-Santiago-Electric, Zhang-Anisotropically, Balberg-Anisotropic} and a CNT-filled silicone 
\cite{Sohn-Seamless} all had larger electrical conductivity in the direction of melt flow than perpendicular to it, which was explained by preferential alignment. In light of this, it is unusual to find that conductivity is larger in the direction normal to shear, as we did here. An important difference to previous studies is the role of the matrix: In two of the three referenced sources on CB-filled thermoplasts, the alignment refers to the polymer chain segments rather than to CB \cite{Vigueras-Santiago-Electric,Zhang-Anisotropically}; electrical anisotropy originated from an anisotropic distribution of CB particles (vs. preferential rotational state of anisometric aggregates and agglomerates) which was imposed by the packing and orientation of the matrix chains during processing (e.g. by CB adsorption to aligned PET fibers \cite{Zhang-Anisotropically}). These examples show that it is incorrect to generally equate flow-induced anisotropy with an alignment of CB particles along the shearing direction. In fact, work on low molecular weight CB suspensions documents anisotropic CB structures aligned perpendicular to the shearing direction \cite{Grenard-Shear,Negi-New,Osuji-Shear,Osuji-Highly,Hipp-PhD}. 

The mechanism leading to flow-induced electrical anisotropy appears to have similar efficiency across the given range of flow kinetics and shear forces. Even the lowest shear rate of 50~s\textsuperscript{-1} is sufficient to induce electrical anisotropy. This is consistent with the fragmentation of a weak network of CB agglomerates upon exceeding a critical shear rate, which is commonly evidenced as shear-thinning behavior \cite{Hipp-Direct, Richards-Review}. Since the latter has been reported for a mixture very similar to our uncured CB-silicone mixtures in the concerned range of shear rates (see \cite{Zhang-Microscopic}: 9 vol\% CB in silicone resin, steady-state viscosity at 50 s\textsuperscript{-1} $\approx$~70 Pa$\cdot$s, vs. $\sim$20 kPa$\cdot$s at 0.1 s\textsuperscript{-1}), we propose that the CB network gets disrupted by the shear forces, becomes anisotropic as a result of shear flow, and reforms rapidly upon cessation of shear.

We will re-visit these structural considerations %, along with a more detailed literature review, 
in Section \ref{Structural analysis}.

\subsection{Piezoresistive anisotropy} \label{Electromechanical properties_results}
Since conductive CB elastomers can be used in sensing applications, we not only consider how flow-induced anisotropy affects electrical conductivity of undeformed films, but how it affects piezoresistivity, i.e., strain-induced resistance changes. For this, we compare the electromechanical response (electrical resistance, mechanical stress) to uniaxial strain parallel vs. perpendicular to the coating direction. Since mechanical anisotropy can lead to piezoresistive anisotropy (see eq.~\eqref{eq:resistance as function of stretch} in Section \ref{Electromechanical properties_experimental}), we first discuss the mechanical stress response to strain (Section \ref{Mechanical stress response to uniaxial strain}). In a second step, the anisotropy of piezoresistive sensitivity is examined (Section \ref{Electrical resistance response to uniaxial strain}).

\subsubsection{Mechanical anisotropy} \label{Mechanical stress response to uniaxial strain}
\begin{figure}
    \centering
    \includegraphics[width=1\linewidth]{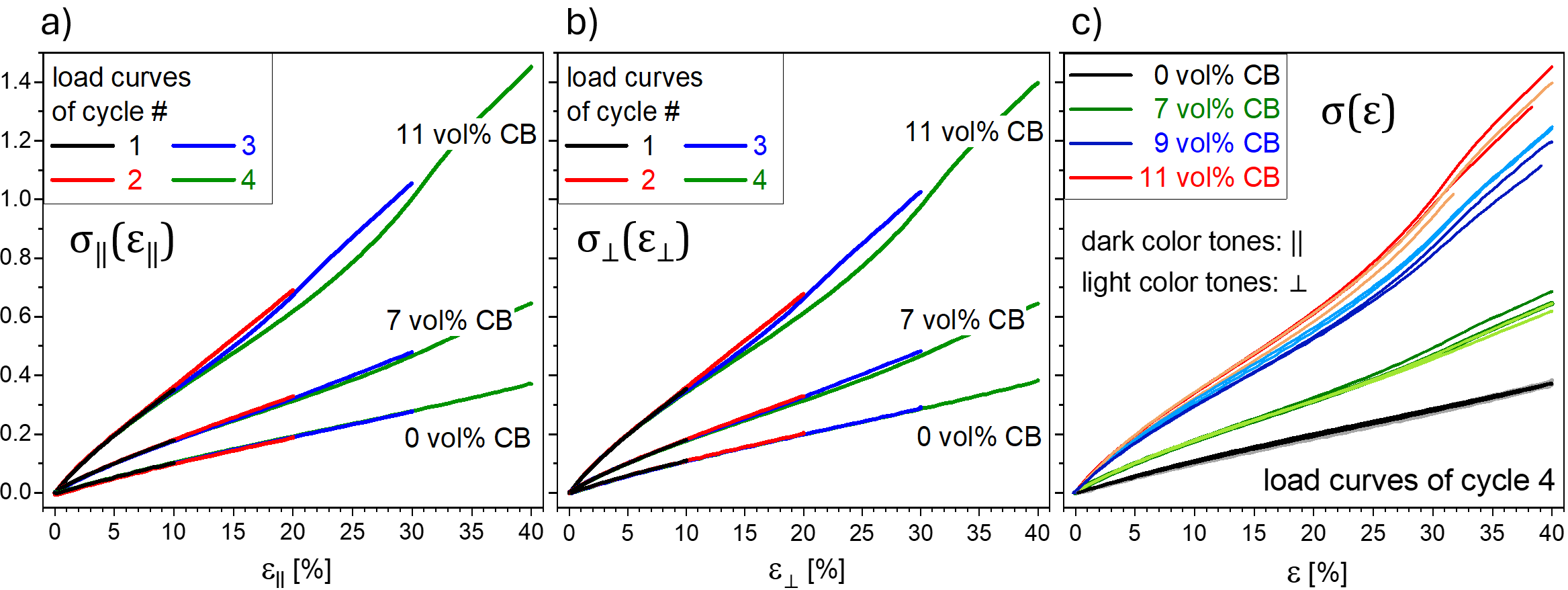}
    \caption{\justifying{The stress response to uniaxial strain (strain rate 10\textsuperscript{-2} s\textsuperscript{-1}, 22°C ± 1 K, 30 ± 15 \% r.h.) of CB-silicone films is isotropic within the limits of accuracy. Shown are the stress-strain curves (nominal stress, $\sigma$, vs. nominal strain, $\epsilon$) of: a) all load cycles for stretch parallel to the coating direction, b) all load cycles for stretch perpendicular to the coating direction, c) load phase of cycle 4 for all samples and stretching directions. In graphs a) + b), the curves for 9 vol\% CB have been omitted for clarity; they lie just below the ones for 11 vol\% CB and have the same shape.}}
    \label{fig:stress-strain curves_parallel&perpendicular stretch}
\end{figure}

The stress-strain curves in Fig. \ref{fig:stress-strain curves_parallel&perpendicular stretch} show that increasing CB concentration reinforced the composites. This is indicative of strong filler-matrix adhesion and a fine dispersion of CB in the matrix (see Section \ref{Nanomechanical mapping (PeakForce QNM)_results} for confirmation). Furthermore, the load cycles for unfilled and CB-filled silicone (Fig. \ref{fig:stress-strain curves_parallel&perpendicular stretch}a-b, 0~vol\% vs. 9/11~vol\% CB) indicate that CB introduced strain softening: The composites became less stiff with increasing cycle number (evident for \(\geq\) 10~\% strain), and the effect was stronger for higher CB contents (maximal for 11~vol\% CB, stress reduction by up to $\sim$15~\%). This is common in filled rubbers and referred to as the Mullins effect \cite{Mullins-Softening}. Microstructural hypotheses for this behavior (see \cite{Huang-Strain} and sources therein, for example) include the rupture of the filler network and the damage of polymer-filler interphases. Irrespective of the mechanism(s) in our specific case, we conclude that stretching induced permanent changes (relaxations and/or microstructural damage) to the CB network and/or the interphase polymer.
    
The tensile tests brought no signs of mechanical anisotropy whatsoever. Figure \ref{fig:stress-strain curves_parallel&perpendicular stretch} illustrates the same stress response for the two stretch axes, $\sigma$\textsubscript{$\parallel$}($\epsilon$\textsubscript{$\parallel$}) and $\sigma$\textsubscript{$\perp$}($\epsilon$\textsubscript{$\perp$}), within experimental scatter of 3~–~5~\% (maximal difference between $\sigma$\textsubscript{$\parallel$} and $\sigma$\textsubscript{$\perp$} relative to $\sigma$\textsubscript{$\parallel$}: 5/5/-6/3~\% for 0/7/9/11 vol\% CB). In contrast, electrical conductivity was clearly anisotropic, both in the unstrained (Section \ref{Local electrical properties (unstrained state)}) and strained state (Section \ref{Electrical resistance response to uniaxial strain}). We conclude that the flow-induced structuration responsible for electrical anisotropy is either irrelevant for the mechanical stress response or its mechanical effect is too weak to be resolved. In the following, we consider the anisotropy of piezoresistance that links mechanical deformation and electrical resistance change.

\subsubsection{Anisotropy of piezoresistive sensitivity} \label{Electrical resistance response to uniaxial strain}
\begin{figure}
     \centering
     \includegraphics[width=0.9\linewidth]{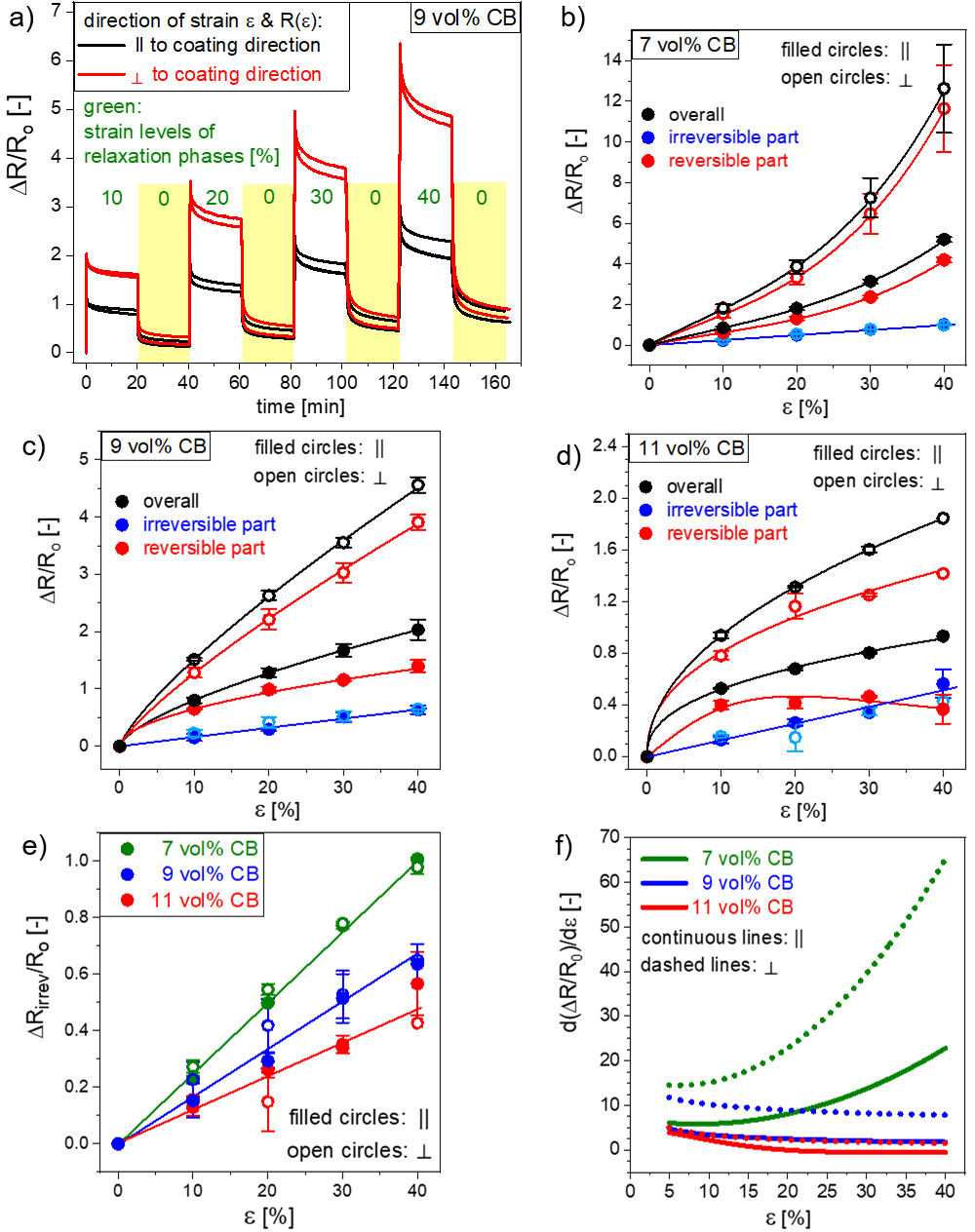}
     \caption{\justifying{Electrical resistance of CB-silicone films (7/9/11 vol\% CB) increased more strongly for strain perpendicular to the coating direction than parallel to it. a) Example of resistance data (9 vol\% CB, change along stretch axis relative to initial value before stretching, $R_{0}$) during load-unload cycles to $\epsilon_{max}$ = 0/10/20/30/40~\% (22°C ± 1~K, 30 ± 15~\% r.h.).
     For quantitative analysis of piezoresistive anisotropy, relaxed resistances are compared for parallel and perpendicular stretch (mean and maximal error from two samples each):  b) – d) overall resistance change relative to $R_{0}$, along with irreversible (permanent increase after unloading) and reversible (overall minus irreversible increase) parts (lines = trend), e) irreversible parts (lines = trend), f) first derivatives of trend curves in panels b-d as a measure of piezoresistive sensitivity.}}
     \label{fig:piezoresistivity compact}
 \end{figure}

Electrical resistances always increased when applying uniaxial strain parallel and perpendicular to the coating direction, respectively. As qualitatively represented by the data for 9~vol\%~CB (Fig. \ref{fig:piezoresistivity compact}a), the resistance increase was much larger for stretch perpendicular to the coating direction than for stretch parallel to it. In conclusion, doctor blade coating the liquid CB-silicone mixtures induced anisotropy in the piezoresistive response, with sensitivity being higher perpendicular to the coating direction (d$R_\perp(\epsilon_{\perp})/$d$\epsilon_{\perp}>$ d$R_\parallel(\epsilon_{\parallel})/$d$\epsilon_{\parallel}$).

To discern influencing factors for flow-induced anisotropic piezoelectric sensitivity and its practical relevance, the next paragraphs address the following questions: 
\begin{itemize}
    \item What are respective contributions of intrinsic resistivity and geometrical changes to the strain-induced resistance increase, and which mechanisms may cause intrinsic resistivity changes?
\end{itemize}
\begin{itemize}
    \item To which extent are strain-induced resistance changes reversible upon unloading?
\end{itemize}
\begin{itemize}
    \item Why do we see clear anisotropy in the electrical response to uniaxial strain but not in the stress response (Section \ref{Mechanical stress response to uniaxial strain})?
\end{itemize}
\begin{itemize}
    \item How do piezoresistive sensitivity and its anisotropy scale with CB concentration?
\end{itemize}
\begin{itemize}
    \item Which implications does the flow-induced (piezo-)electric anisotropy have for sensing applications?
\end{itemize}

In general, both intrinsic electrical resistivity and sample geometry contribute to resistance. 
As explained in Section \ref{Electromechanical properties_experimental}, we cannot discriminate them due to the unknown (possibly anisotropic) strain-dependence of the cross-sectional area. 
In light of the continuity of the resistance increase (Fig. \ref{fig:piezoresistivity compact}b-d), we assume that both contributions grow upon stretching, at least up to our moderate maximal strains of 40~\%. 
Concerning intrinsic resistivity, a strain-induced increase is commonly observed for CB elastomers at small to moderate stretches up to 25~-~30~\% \cite{Flandin-Effect,Knite-Polyisoprene,Knite-Reversible,Zhang-Anisotropically,Yamaguchi-Electrical}. 
The increase is attributed to growing CB interparticle distances, i.e., conductive pathways become more resistive and fewer in number when tunneling distances increase and electrical contacts are lost as a result of stretching \cite{Kost-Resistivity,Knite-Reversible,Wang-Highly,Zhang-Anisotropically,Yamaguchi-Electrical}. 
This agrees with structural evidence for a CB elastomer presented in \cite{Ehrburger-Dolle-Anisotropic} where the degree of interpenetration of CB aggregates diminishes parallel to the stretch axis. Concerning the structural origin of piezoresistive anisotropy, we have not found any work in literature. In light of the isotropic stress response, we assume that the geometrical contribution  to $R(\epsilon)$ is irrelevant for piezoresistive anisotropy, and that the latter is solely based on anisotropic changes in intrinsic resistivity. We point out that a net increase of interparticle distance in the stretching direction is not necessarily the governing mechanism at small to moderate strains: The strain-induced translation and rotation of filler particles generally leads to both the breakdown and formation of conductive pathways, and either of the two may dominate during stretching. For example, Flandin et al. \cite{Flandin-Effect} report that for a high structure CB-silicone composite, resistance decreases for  up to 5 \% strain and attribute this to the net creation of new conduction paths or improvement of existing ones.
In addition to increasing interparticle distances, particle alignment along the stretch axis is reported to become significant at moderate strains for various conductive CB elastomers (e.g. for strains $>$ 30 \% \cite{Yamaguchi-Electrical}, $>$ 30 \% \cite{Kost-Effects}, $>$ 25 \% \cite{Kost-Resistivity}). We will discuss this mechanism in Section \ref{Structural analysis} for our films.

To discriminate reversible and irreversible effects, Fig. \ref{fig:piezoresistivity compact}b-e displays the relaxed resistance increase at the probed strain plateaus (derived from $R(t)$ as explained in Appendix \ref{appendix: electromechanical}) relative to resistance of unstretched samples, $R_0$, along with its irreversible (resistance change after unloading from a given strain plateau and relaxing) and reversible (overall minus irreversible increase) parts. As seen in Fig. \ref{fig:piezoresistivity compact}e, the irreversible part increases quasi-linearly with strain. It reaches values of $\geq$ 1/10 $R_0$ at the smallest strain plateau ($\epsilon_{max}$ = 10 \%), and up to 1/2 to 100 \% of $R_0$ at the biggest ($\epsilon_{max}$ = 40 \%). We mainly attribute the irreversible resistance increase to intrinsic resistivity changes (i.e., permanent modifications of the CB network) rather than geometric ones (i.e., plastic deformation at the macroscale). As explained in Appendix \ref{appendix: electromechanical}, plastic strains amount to 4~\%, which would account for a resistance increase by the same, small percentage (via the sample length $L$, see eq. \eqref{eq:resistance as function of stretch} in Section \ref{Electromechanical properties_experimental}). A significant permanent decrease in the cross-sectional area is also excluded as this would seriously raise the nominal stress values (see Section \ref{Mechanical stress response to uniaxial strain}, Fig. \ref{fig:stress-strain curves_parallel&perpendicular stretch}). As a cause for permanent resistivity increase, literature frequently considers the damage of filler-matrix interphases (also termed ‘debonding’ or ‘slippage’ at the filler-matrix interface) \cite{Kost-Resistivity,Yamaguchi-Electrical,Knite-Electric}. Mechanical failure at the interphases is plausible here since the stress response is also indicative of it (strain softening for strains $\geq$ 10 \%, see Section \ref{Mechanical stress response to uniaxial strain}); however, the exact role of filler-matrix delamination for electrical conductivity remains obscure. For example, Yamaguchi et al. \cite{Yamaguchi-Electrical} argue that the effect of filler-matrix delamination on the spatial arrangement of filler particles should be rather small, whereas Knite et al. \cite{Knite-Electric} hypothesize the irreversible formation of isolated filler clusters. Irrespective of the exact mechanism, we conclude that straining provokes microstructural changes in the composite that result in a different set of conductive pathways of the CB network after unloading. Interestingly, the irreversible increase of resistance along the stretch axis is isotropic (Fig. \ref{fig:piezoresistivity compact}b-e), as opposed to electrical resistance in the unstrained state (Section \ref{Local electrical properties (unstrained state)}) and the reversible resistance increase (Fig. \ref{fig:piezoresistivity compact}b-d). In conclusion, only the reversible part of the resistance response to strain caused piezoresistive anisotropy. 

As a reminder, the stress response is isotropic within an experimental scatter of 3~–~5~\% (Section \ref{Mechanical stress response to uniaxial strain}). In contrast, anisotropy in $R(\epsilon)$ is clearly significant, even within the more pronounced scatter of $\leq$~17~\% (e.g. relative difference in resistance increases for stretch parallel vs. perpendicular to the coating direction, $\Delta R(\epsilon_\parallel)/R_0$ and $\Delta R(\epsilon_\perp)/R_0$ at 40 \% strain: 243/225/198~\% for 7/9/11 vol\% CB). Apparently, electrical properties of the composite are much more strongly affected by flow-induced anisotropy than mechanical properties. A basic reason for this is that electrical conductivity necessitates the existence of percolated paths of CB, whereas the reinforcing effect of CB does not. Concomitantly, small changes of the CB network can account for significant changes in conductivity (via loss/gain of conductive pathways) without (or barely) affecting mechanical reinforcement – a phenomenology that has also been reported in \cite{Yamaguchi-Electrical}.

We now consider the effect of CB concentration on piezoresistive sensitivity and its anisotropy, as mirrored in Fig. \ref{fig:piezoresistivity compact}b-f. The reversible (and irreversible) part of $\Delta R(\epsilon)$ gets stronger upon lowering CB concentration from 11 vol\% to 7 vol\%, i.e., piezoresistive sensitivity increases upon approaching the percolation threshold ($\sim$5 vol\% \cite{Zhang-Microscopic}). The trend is commonly noted for deformation-induced conductivity changes in CB composites \cite{Karuthedath-Characterization,Kost-Effects,Kost-Resistivity,Knite-Reversible}. As a general explanation in terms of percolation, conductivity is most sensitive to microstructural changes when the filler concentration barely suffices to form a few conductive paths, and the loss of these paths cannot be compensated by a recombination with particles from the vicinity. 
%More experimental and simulative work is needed for a detailed understanding, both for irreversible effects involving non-equilibrium processes (e.g. filler-matrix delamination) and for the reversible resistance increase which may involve multiple mechanisms. 
This picture is supported by our data since the resistance increase as a function of strain transitions is convex close to the percolation threshold (7~vol\% CB): A convex dependence is typical of only a few single conductive paths which can be modeled as conductive elements connected in series. In contrast, the dependence is concave for higher CB concentrations (9/11~vol\% CB), characteristic of particles in a denser network which are increasingly connected in parallel. 
%This points to a shift in underlying mechanisms or their hierarchy, e.g. the relative relevance of contact and tunneling conduction. 
Concerning the anisotropy of piezoresistive sensitivity, it is also strongest when closest to the percolation threshold (compare relative differences between parallel and perpendicular stretch for the three CB contents in Fig. \ref{fig:piezoresistivity compact}f). This is analogous to observations on anisotropy in carbon-filled polymers, e.g., for flow-induced electrical anisotropy of CB thermoplasts \cite{Balberg-Anisotropic,Zhang-Anisotropically}, electrical anisotropy of magnetically aligned graphite fibers in an epoxy \cite{Carmona-Anisotropic}, and anisotropic X-ray scattering of the CB network in ethylene propylene rubber \cite{Ehrburger-Dolle-Anisotropic}. The trend of electrical anisotropy being maximal just above percolation is also seen for a simple model for anisometric conductive particles aligned in a dielectric continuum \cite{Carmona-Anisotropic}, i.e., it can be derived from geometrical considerations in terms of preferential filler orientation. It is open to question whether this simple structural idea is adequate in the case of our films and other anisotropic CB polymers, already because there are anisotropic CB structures that are not governed by aggregate alignment (see review in Section \ref{Structural analysis}). In addition, the impact of filler concentration on hydrodynamics during processing must be taken into account (Section \ref{Local electrical properties (unstrained state)}): Structuration mechanisms leading to anisotropy may become less efficient with increasing CB concentration as the material gets more viscous and resistant to flow.

The above phenomenology has important implications for the application of CB elastomers and other composites with conductive fillers. Our results show that electrical anisotropy from shear flow during film fabrication may be insignificant in the undeformed state but become pronounced upon deformation. As an example from our data, initial anisotropy is fairly weak with mean values of $R_\parallel/R_\perp$ = 1.1 – 1.4, see Section \ref{Local electrical properties (unstrained state)}), whereas at 40 \% strain, $R_\parallel/R_\perp$ is as low as 0.5, i.e., the material is twice as resistant along the stretch axis when stretched perpendicular to the coating direction as when stretched parallel to it. Thus, in cases where (piezo-)electric anisotropy is not desired, one should test for it not only in the undeformed state but also along at least two stretch axes at strains relevant in practice. As a further practical measure, our data indicate that CB concentration should be chosen well above the percolation threshold since anisotropic effects in conductivity and piezoelectric sensitivity are then weaker. One could, however, also think of applications where (piezo-)electric anisotropy is favorable, and tune CB elastomers toward maximal piezoelectric anisotropy. For the interested reader, we therefore report on the anisotropic sensing performance of our CB-silicone films in Appendix \ref{Sensing performance}.

\subsection{Structural analysis} \label{Structural analysis}

Before presenting our results from structural characterization, we compile some general considerations relevant for identifying the mechanisms that lead to flow-induced anisotropic CB structures during doctor blade coating. As mentioned in Section \ref{introduction}, there seems to be no systematic work on flow-induced anisotropy in CB-filled polymers, as opposed to CB suspensions in low viscosity organic liquids \cite{Grenard-Shear,Negi-New,Osuji-Shear,Osuji-Highly} and other colloidal systems with aggregating particles (e.g. \cite{Hoekstra-Multi,Eberle-Shear,Varadan-Shear-induced}). %We therefore include a review  of the referenced publications, in order to get an idea on possible flow-induced anisotropic CB morphologies in CB elastomers.
\begin{itemize}
    \item As a basic requirement for flow-induced anisotropy, shear forces have to be strong enough to fragment the CB network. The latter manifests as shear-thinning flow in rheometry \cite{Hipp-Direct, Richards-Review}. Since the global shear rates of our coating process fall deep into the shear-thinning regime of a similar CB-silicone mixture (see \cite{Zhang-Microscopic} and Section \ref{Local electrical properties (unstrained state)}), we expect doctor blade coating to be highly effective in breaking up the CB network and allowing for translation and rotation of CB agglomerates and primary aggregates.
\end{itemize}

\begin{itemize}
    \item Work on CB in low viscosity suspending media reports that for high shear rates ($10^2$ – $10^3$~$s^{-1}$ for CB in tetradecane and CB in mineral oil \cite{Osuji-Highly,Osuji-Shear,Negi-New}), primary aggregates are broken, resulting in reduced sizes and manifesting as shear-thickening flow. Even though our global shear rates (50~$s^{-1}$ to 3000~$s^{-1}$) cover this regime and would equate to even higher shear forces (due to the higher matrix viscosity), we do not expect a significant break-up of primary aggregates in light of the shear-thinning of a similar CB-silicone mixture in the concerned range of shear rates (see preceding point). To verify, we derive an aggregate size distribution from PeakForce QNM maps and compare it with reference data published previously (Section \ref{Segmentation and statistical analysis of nanomechanical data_results}).
\end{itemize}

\begin{itemize}
    \item A peculiarity of polymeric matrices (vs. low molecular weight matrices) is that they themselves can become anisotropic as a result of shear. As reviewed in Section 3.1, electrical anisotropy can stem from an anisotropic CB dispersion imposed by the packing and orientation of matrix chains during processing \cite{Vigueras-Santiago-Electric,Zhang-Anisotropically}. The publications refer to semicrystalline thermoplasts, i.e., polymers where preferential chain orientation seems more likely than for our silicone matrix which is chemically crosslinked (steric hindrance to chain orientation) as well as amorphous during processing (coating at room temperature, cure at 100 °C) and characterization at room temperature \cite{Brounstein-Longterm} (no tendency for alignment via crystallization). Yet, since we cannot generally exclude anisotropic effects, structural analysis must check for matrix anisotropy. This is done via SAXS (Section \ref{SAXS_results}).
\end{itemize}

\begin{itemize}
   \item Despite the complexity and diversity of flow-induced CB morphologies, one anisotropic version is reported both for CB suspensions [34-37] and short-range attraction colloidal suspensions of other particles (e.g. \cite{Eberle-Shear,Hoekstra-Multi}): Under certain shear flow conditions (e.g. shear rate, filler concentration and size of the shear gap), vorticity-aligned structures (rods or sheets) form parallel to the compressional and neutral axes of shear flow.  %The basic structural idea, applied to the parallel plate situation and resulting films of our coating process, is depicted in Fig. \ref{fig:simple shear flow schematic} (neutral direction = 2-direction in the vorticity plane corresponding to the lateral dimension of the films). The vorticity-aligned structures are large-scale, with widths as high as $10^2$ µm \cite{Osuji-Highly,Grenard-Shear}. 
   As the underlying mechanism, literature suggests the break-up of agglomerates and subsequent separation of aggregates/agglomerates along the extensional axis of shear flow, in combination with the densification/interpenetration of CB aggregates along the compressional axis \cite{Eberle-Shear,Osuji-Highly,Negi-New,Hoekstra-Multi,Vermant-Flow}.

   % \begin{figure}
    %    \centering
   %     \includegraphics[width=0.85\linewidth]{doctor blading simple shear flow schematic2.png}
   %     \caption{\justifying{As a possible model of anisotropy caused by shear flow during doctor blade coating, we consider vorticity-aligned CB structures as reported in literature (see text): Particles tend to be separated along the axis of maximal extension and densified along the axis of maximal compression as visualized in panel a), leading to ‘rows’ of particles densified parallel to the compressional axis (see flow gradient plane in panel b) and perpendicular to the coating direction (see vorticity plane in panel b).}}
  %      \label{fig:simple shear flow schematic}
  %  \end{figure}

In light of their generic appearance in the referenced systems, we shall consider %vorticity-aligned structures for our films. If present, they will forcibly be less pronounced than in the schematic of Fig. \ref{fig:simple shear flow schematic}b, as this ideal periodic 2D-arrangement is not percolated in the 1-direction, and would result in pronounced mechanical anisotropy. A more realistic structural picture is%
that CB aggregates may become separated along the axis of maximal extension of shear flow / parallel to the coating direction. Concerning the undeformed state, this would readily explain why parallel resistance is higher than perpendicular resistance. 
    
\end{itemize}

\subsubsection{Nanomechanical mapping (PeakForce QNM)} \label{Nanomechanical mapping (PeakForce QNM)_results}

The dispersion state of near-surface CB in unstretched samples is illustrated by the 20x20 µm$^2$ dissipation maps in Fig. \ref{fig:PF-QNM comparison CB7-9-11, CB7 strained-unstrained}a-c (for explanation and qualitative discussion of the Peak Force QNM signal maps, see Appendix \ref{Appendix: PeakForce QNM maps and value distributions}). The area density of near-surface CB markedly grows with CB concentration, and CB is finely dispersed in the matrix for all compositions (aggregate and agglomerate sizes of some 100 nm, see Fig. \ref{fig:PF-QNM comparison CB7-9-11, CB7 strained-unstrained}d-e as well as Section \ref{Segmentation and statistical analysis of nanomechanical data_results} for segmental analysis). Upon assuming that both observations hold true also for the bulk, we have proven the structural basis for the mechanical stiffening by CB (Section \ref{Mechanical stress response to uniaxial strain}): The rather homogeneous dispersion of 100 nm-sized aggregates provides a sufficient portion of reinforcing CB-matrix interphases, and the latter become more prominent in the composite with increasing CB content. Note that the maps (Fig. \ref{fig:PF-QNM comparison CB7-9-11, CB7 strained-unstrained}, plus more examples in Fig. \ref{fig:threshold}, Section \ref{Segmentation and statistical analysis of PeakForce QNM data_experimental} as well as Fig. \ref{fig:PF-QNM_all signals overview} and Fig. \ref{fig:PF-QNM signal maps typical scaling}, Appendix \ref{Appendix: PeakForce QNM maps and value distributions}) show hardly any agglomerates, and that they appear to be no bigger than 1 µm. In contrast to primary aggregates, agglomerates may not be fully captured as their size may exceed the information depth (some 10\textsuperscript{1} nm to 10\textsuperscript{2} nm). As a result, PeakForce QNM probably underestimates agglomerate size.

\begin{figure}
    \centering
    \includegraphics[width=1\linewidth]{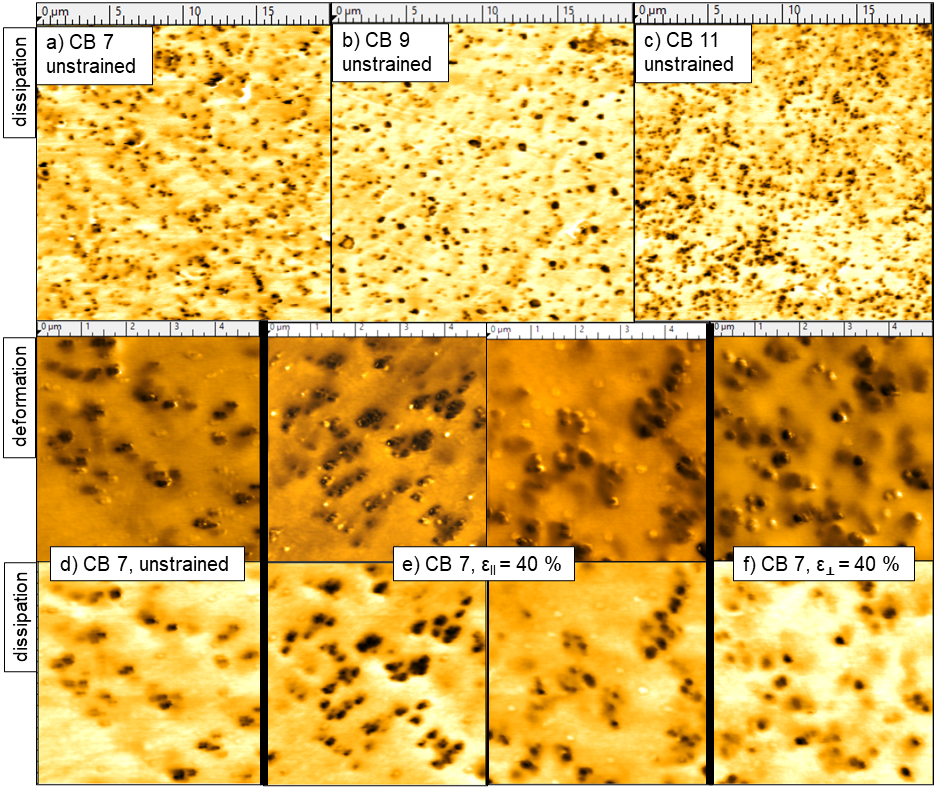}
    \caption{\justifying{PeakForce QNM maps of the CB-silicone films reveal that the near-surface CB morphology is uniform for all examined CB contents (7/9/11 vol\% CB, unstretched, 20x20 µm\textsuperscript{2} scans, panels a-c). The impact of uniaxial strain on CB particle distance and orientation is not readily discernible via qualitative inspection of the data, as seen by the deformation and dissipation maps of 5x5 µm$^2$ measuring spots of 7~vol\% CB samples in the d) unstrained state vs. e-f) strained, relaxed states at $\epsilon_\parallel$ = 40~\% and $\epsilon_\perp$ = 40~\%, respectively. In each of the maps, the coating direction goes from bottom left to top right.}}
    \label{fig:PF-QNM comparison CB7-9-11, CB7 strained-unstrained}
\end{figure}

Concerning anisotropy and strain-induced changes, the 20x20 µm$^2$ scans do not provide sufficient resolution to discern possible preferential orientation of CB aggregates and agglomerates. Instead, we turn to the more highly resolved 5x5 µm$^2$ maps in Fig. \ref{fig:PF-QNM comparison CB7-9-11, CB7 strained-unstrained}d-f. Shown are different scan areas for both unstrained and strained  states of the 7 vol\% CB film, i.e., the composition with the strongest piezoelectric effects (see Section \ref{Electrical resistance response to uniaxial strain}). The signal maps indicate that a significant portion of CB aggregates (visible as coherent dark spots) and agglomerates (clusters of aggregates separated by thin yellowish lines in the maps) is anisometric, i.e., the basic condition for their preferential orientation is met. Yet, the exemplary maps for the unstrained state (Fig. \ref{fig:PF-QNM comparison CB7-9-11, CB7 strained-unstrained}d) and perpendicular strain (Fig. \ref{fig:PF-QNM comparison CB7-9-11, CB7 strained-unstrained}f) do not readily evidence preferential alignment. Some smaller aggregates appear oriented horizontally, but this effect is not reliable as we cannot exclude scanning bias from the height control of the SFM cantilever. (It is for this reason that we have chosen a scanning angle of 45°, i.e., the horizontal bias will have equal effects in the two stretching directions.) The maps for the state strained parallel to the coating direction give an ambivalent picture with regard to particle alignment: In some regions, aggregates appear preferentially oriented along the axis of coating/stretching (left part of Fig. \ref{fig:PF-QNM comparison CB7-9-11, CB7 strained-unstrained}e), while they tend to be aligned perpendicular to the axis in other regions (right part of Fig. \ref{fig:PF-QNM comparison CB7-9-11, CB7 strained-unstrained}e). 

In conclusion, visual inspection of PeakForce QNM data does not allow statements on changes in CB morphology induced by doctor blade coating or stretching. We therefore opt for a more precise and comprehensive means of data interpretation, i.e., segmentation of a bigger overall data set for the three material states. Corresponding results are presented in the next section. We point out that concerning strain-induced damage, the lateral scan resolution does not suffice to detect voids smaller than $10^1$ – $10^2$~nm. We can, however, exclude the formation of bigger, microscopic pores, since the maps do not indicate any qualitative differences in material contrast between unstretched and stretched states (see Fig. \ref{fig:PF-QNM comparison CB7-9-11, CB7 strained-unstrained} as well as Fig. \ref{fig:threshold} in Section \ref{Segmentation and statistical analysis of PeakForce QNM data_experimental}). The only hint of defects are a few brighter spots of some $10^1$~nm in the matrix-dominated regions of the deformation signal, which could reflect pores in the silicone. However, they are also visible in the unstrained state, and the evidence is too weak overall for an unambiguous assignment.

\subsubsection{Segmentation and statistical analysis of nanomechanical data} \label{Segmentation and statistical analysis of nanomechanical data_results}

\begin{figure}
     \centering
    \includegraphics[width=1\linewidth]{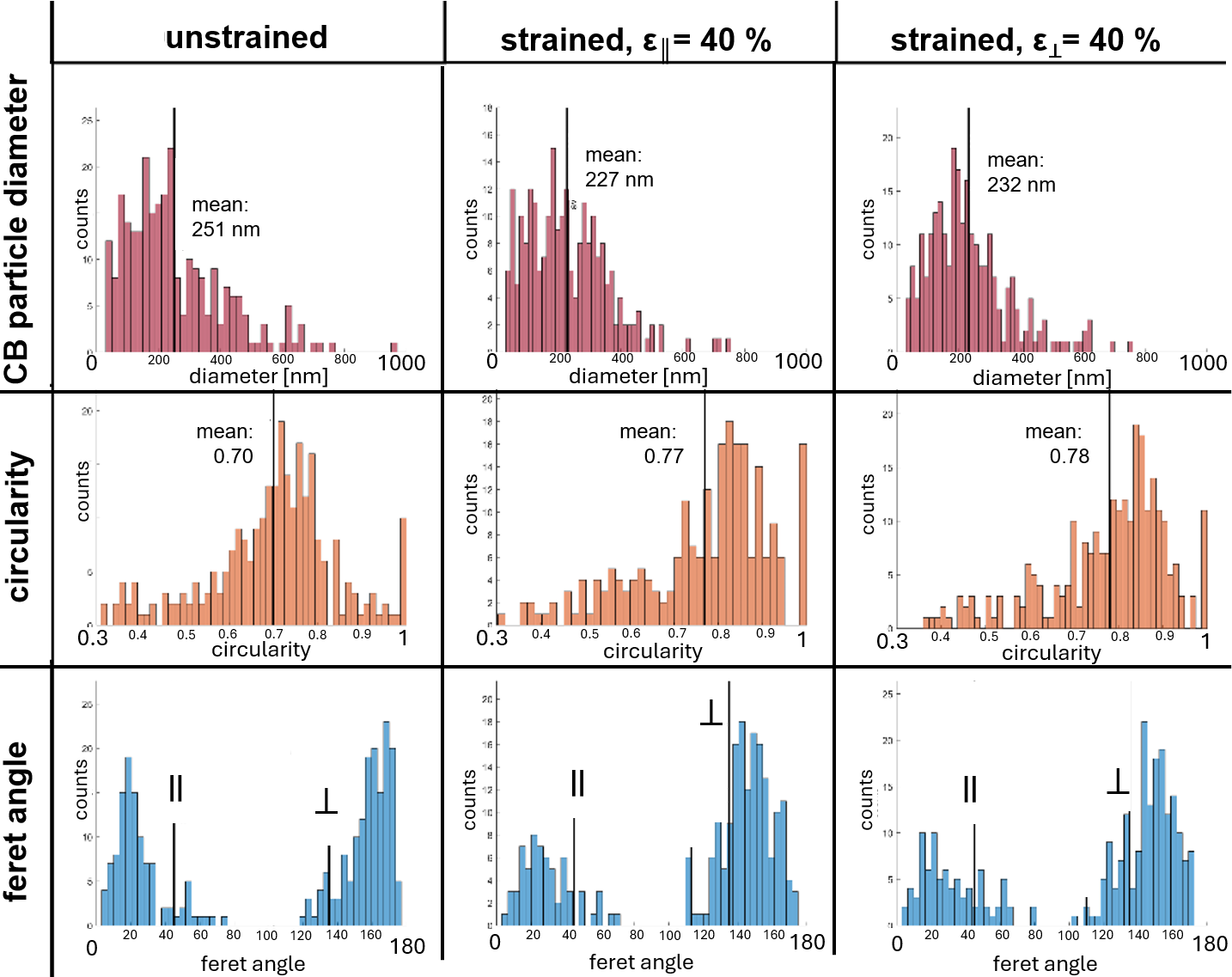}
     \caption{\justifying{Assessment of CB particle diameter, circularity and angle of the major feret axis to the horizontal axis of the PeakForce QNM scans by histograms obtained from segmentation analysis of 10x10 µm$^2$ dissipation maps. Particles are anisometric (circularity $<$ 1), and their size roughly follows a log-normal distribution in the range of about 40 nm (primary diameter) up to about 1~µm. The feret angle reflects weak evidence of two preferential orientations which tend to align with the coating direction ($\parallel$-sign in the histograms) and perpendicular to the coating direction ($\perp$-sign in the histograms) upon straining.}}
    % \label{fig:results}
    \label{fig:segmentation results_edit BZ}
 \end{figure} 

The CB particle distribution was assessed after segmentation of the 10 x 10 µm$^2$ dissipation maps. Three maps were evaluated for each material state (unstrained, strained to 40~\% parallel and perpendicular to the coating direction). The resulting histograms and average values of the particle diameter, circularity as well as the angle of the major feret axis to the horizontal direction are shown in Fig. \ref{fig:segmentation results_edit BZ}. %The average particle fraction after erosion and dilatation agrees well with the preset volume fraction of 7 vol\% and reflects a potential sedimentation by a slight overestimation. 

%The average particle diameter is around 250 nm with a slight tendency to reduce by straining. The particle diameter distributions that reflect roughly a log-normal distribution give no starting point to distinguish between aggregates and agglomerates since the pixel size is about 11 nm and the primary particles are expected to have a diameter between 1.5 and 20 µm$^2$. The same holds true for the distance of particle to their nearest neighbor. Just a slight tendency to reduced the particle distance as a result of straining was observed. 

The CB particle diameter roughly follows a log-normal distribution with values in the $10^2$~nm-range, confirming the size range derived from visual inspection of the PF-QNM maps in the previous section. In addition, the distribution agrees very well with aggregate size measurements for CB suspended in toluene \cite{Coupette-Percolation}, indicating that (as expected) primary aggregates are not modified / broken down by shear during the coating process. (Note that the histograms provide no means to differentiate between primary aggregates and agglomerates; the same was found for USAXS measurements on bulk composites (3/5/7/9 vol\% CB) where scattering by both species strongly overlapped \cite{Coupette-Percolation}.)

%The circularity parameter, on the other hand, responds significantly to straining. The particles appear to become rounder when the sample is stretched, regardless of whether the stretching is parallel or perpendicular to the coating direction. 

The circularity parameter is significantly smaller than 1 (means: 0.7 - 0.8), confirming the visual particle assessment in the previous section, i.e., CB aggregates are anisometric and can, as a result, exhibit preferential orientation.

%The feret angle shows a clear tendency to fluctuatearound the horizontal axis. This can be attributed to the smearing of the particle measurement by the SPM technique described above, as the scanning direction is horizontal. 

The histogram of the feret angle for the unstrained film does indeed show a trend to two maxima at roughly 20° and 170° relative to the horizontal (scanning) direction. This can be an indication of a slight preferential orientation; however, it is superimposed by a SFM scan artifact (smearing of particles in the horizontal/ 0°-direction) which prevents a definite conlusion. Since the coating direction is at 45° in the scans, preferential particle alignment from the coating process  or straining would rather be expected at the 45° or 135° direction. %It should be noted that in the unstretched state, 0° or 180° do not coincide with the maximum in the histogram, but there exist deviations from this axis by about 10°. 
 With straining, the maxima of the histogram appear to shift from the extremes at approx. 20° and 170° towards the coating direction and perpendicular to it. This indicates a slight alignment of the CB particles due to the straining, but just like for the unstrained state, the evidence is too weak for an unambiguous interpretation.

%o	if no apparent differences to results in \cite{Coupette-Percolation}: aggregates not modified/broken up by shear

%Florian: Ich sehe in den 5x5 Bildern zu wenige Partikel, insbesondere, wenn ich die Randpartikel sinnvollerweise aus der Auswertung herausnehme und detektiere nur verrauschtere Histogramme. Ich kann leider so die Aggregate, Primärpartikel und Aggregate nicht auftrennen. Es wird zwar in den Histogrammen teilweise so, als wäre da um die mittlere Partikelgröße eine kleine Delle, aber das wäre mir zu windig, um das aus den verrauschten Daten herauszulesen. Ich würde diese Einschätzung Dir als Expertin überlassen. 

\subsubsection{SAXS} \label{SAXS_results}

Small-angle X-ray scattering (SAXS) measurements were performed to study the bulk structure of unstretched and stretched silicone films, both neat and filled with CB at a concentration of 9~vol\%. 

\begin{figure}
    \centering
    \includegraphics[width=1.0\linewidth]{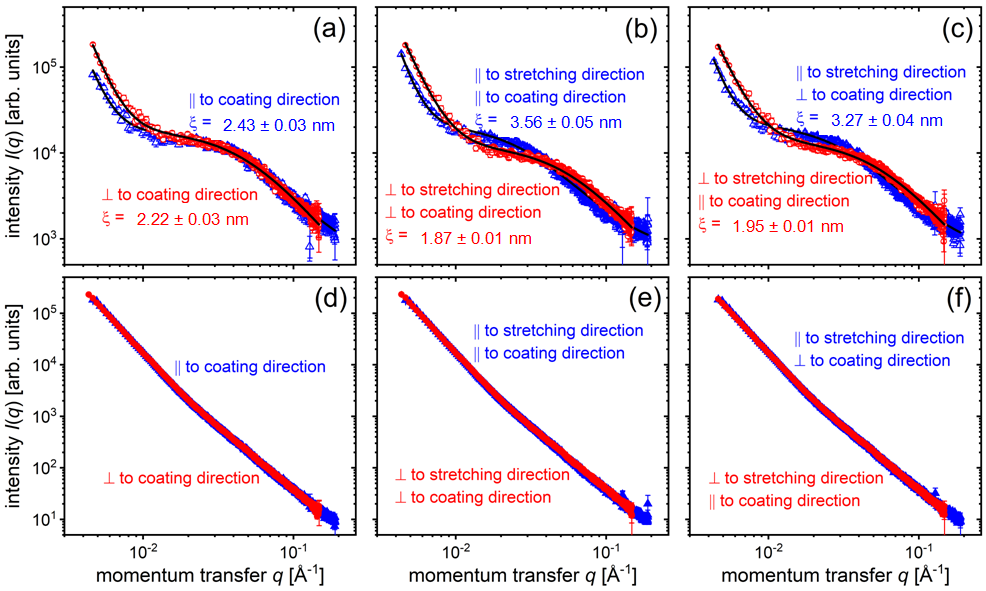}
    \caption{\justifying{1D SAXS patterns of unloaded silicone films (a,b,c) and silicone films loaded with 9~vol\% CB (d,e,f) (h\textsubscript{gap} = 350 µm, v\textsubscript{blade} = 20 mm/s) in their unstretched state (a,d), stretched at 40 \% uniaxial strain parallel to the coating direction (b,e) and perpendicular to the coating direction (c,f). The patterns are evaluated in the directions as indicated in the graphs, detailed in the experimental section. The black lines are fits according to eq.~\eqref{eq:saxsfit}.}}
    \label{fig:SAXS curve_CB9 unstrained vs strained}
\end{figure}

Fig. \ref{fig:SAXS curve_CB9 unstrained vs strained}a presents 1D scattering patterns of the neat silicone film in its unstretched state, obtained by azimuthally averaging the 2D scattering patterns in a narrow angular range, both in the coating direction and perpendicular to it (see experimental section for more information). Both curves feature a shoulder at $\sim$0.04 \AA$^{-1}$, assigned to scattering of the silicone network, and weak forward scattering due to large-scale structures. The latter may be assigned to frozen-in elastic forces inside the gel, leading to large-scale composition fluctuations \cite{geissler1997, urayama1998}. To quantify the involved length scales, the scattering curves were modeled using a contribution accounting for large-scale structures, $I_{\text{LS}}(q)$, and a contribution accounting for composition fluctuations, $I_{\text{fluct}}(q)$, following the equation
\begin{equation} \label{eq:saxsfit}
I(q) = I_{\text{LS}}(q) + I_{\text{fluct}}(q) + I_{\text{bkg}} 	
\end{equation}
with $I_{\text{bkg}}$ an constant background. A generalized Porod law is used to account for scattering by the large scale structures, and is given by
\begin{equation} \label{eq:saxsPorod}
I_{\text{LS}}(q) = \frac{K_{\text{P}}}{q^{m}} 	
\end{equation}
with $K_{\text{P}}$ the Porod amplitude and $m$ the Porod exponent. The Ornstein-Zernike structure factor accounts for scattering by composition fluctuations on the level of single chains. It is given by
\begin{equation} \label{eq:saxsOZ}
I_{\text{fluct}}(q) = \frac{I_{\text{OZ}}}{1+\xi^2 q^2}	
\end{equation}
with $I_{\text{OZ}}$ the Ornstein-Zernike amplitude and $\xi$ the correlation length of composition fluctuations \cite{geissler1997, urayama1998}.

$\xi$, which is proportional to the mesh size of the silicone network and thus a measure of crosslink density, equals $2.43 \pm 0.03$ nm parallel to the coating direction and $2.22 \pm 0.03$ nm perpendicular to the coating direction. Thus, the coating process introduced a notable structural anisotropy at the nanoscale of the silicone network, presumably via increased flow of chain segments along the extensional axis of shear (see Fig. \ref{fig:tensile & 4PP specimens, 4PP polarity, planes+directions}c). In addition, the scattering at large-scale structures ($q < 10^{-2}$) is significantly stronger perpendicular to the coating direction, implying that they are larger in size and/or number. Data for yet lower $q$-values would be necessary to draw conclusions on possible anisotropy at these larger length scales of some $10^2$ nm.

Strain significantly changed the nanoscale structure of the silicone network, as is shown in Fig. \ref{fig:SAXS curve_CB9 unstrained vs strained})b for 40 \% uniaxial strain applied parallel to the coating direction and Fig. \ref{fig:SAXS curve_CB9 unstrained vs strained})c for 40 \% uniaxial strain applied perpendicular to the coating direction. In the direction of strain, the mesh size is significantly larger than perpendicular to the direction of strain (see $\xi$-values in Fig. \ref{fig:SAXS curve_CB9 unstrained vs strained}b-c), reflecting transverse contraction. To quantify the corresponding compressibility, the Poisson ratio for finite strains according to Hencky, $\nu = -ln(\lambda_{trans})/ln(\lambda_{axial})$, was applied to the $\xi$-values from the SAXS fits. The extensional ratios in the transverse and axial directions, $\lambda_{trans}$ and $\lambda_{axial}$, are thus each derived as the quotient of the $\xi$-values for the strained vs. unstrained state. The resulting values for stretch parallel vs. perpendicular to the coating direction, $\nu_\parallel$ = 0.45 $\pm$ 0.09 and $\nu_\perp$ = 0.44 $\pm$ 0.11, are both equal to 0.5 within the relative uncertainty of 20 - 25 \%, indicating nearly to fully incompressible behavior ($\nu = 0.5$), typical for rubbers. In light of the slightly anisotropic mesh size reported above (greater in the parallel direction than in the perpendicular direction by 9 \%), this either means that small directional differences in crosslink density do not lead to an anisotropic compressibility, or that their effect on compressibility is too weak to resolve here. By all means, the nanoscopically derived, isotropic Poisson ratios agree well with the macroscopic, also isotropic stress response reported in Section \ref{Mechanical stress response to uniaxial strain}.

%Therefore, the deformation is affine, i.e., the applied macroscopic strain is transferred uniformly to the individual meshes. 

% Hier ist ein Entwurf, falls man auf Makro- vs. Mikrodehnung eingehen will: 
%To further substantiate the agreement of microscopic and macroscopic deformation, we note that the microscopically derived strains based on $\xi$, $\epsilon_\parallel = 46.5 \%$ and $\epsilon_\perp = 34.6 \%$, are close to the macroscopically applied nominal strain of 40 \% (46.5 \% and , respectively). 

%Furthermore, in both cases, the mesh sizes parallel to the coating direction were slightly larger than those perpendicular to the coating direction, consistent with the mesh sizes of the unstretched silicone film.

The scattering patterns of the silicone films loaded with CB at a concentration of 9 vol\% (Fig. \ref{fig:SAXS curve_CB9 unstrained vs strained})d) show a steep slope in the entire measured $q$ range, indicating that structures (primary CB particles and their aggregates with a very broad size distribution) are present at all length scales covered in the measurement. The scattering is dominated by CB and the contrast of the silicone network is too weak to resolve. No difference in the scattering patterns evaluated parallel and perpendicular to the coating direction is observed. Therefore, coating does not introduce structural anisotropy on length scales between $\sim$1 and $\sim$150 nm, i.e., to CB primary particles (about 40 nm \cite{Coupette-Percolation}) and small aggregates.

Also stretching the 9 vol\%-composite, both parallel (Fig. \ref{fig:SAXS curve_CB9 unstrained vs strained})e) and perpendicular (Fig. \ref{fig:SAXS curve_CB9 unstrained vs strained})f) to the coating direction, did not lead to any structural anisotropy on length scales between $\sim$1 and $\sim$150 nm. In both cases, the scattering curves evaluated parallel and perpendicular to the direction of stretching did not deviate from each other.

%→ matrix compressibility isotropic, consistent with mechanical isotropy (Section \ref{Mechanical stress response to uniaxial strain})

%•	no anisotropy according to SAXS for structures $\leq$ 150 nm: might not be able to resolve weak orientational effects on CB aggregate level, since portion of aggregates $\leq$ 150 nm very small (see Section \ref{Segmentation and statistical analysis of nanomechanical data_results})? Would mean that anisotropy on aggregate level cannot be safely excluded 

\subsection{Simulations} \label{Simulations_results}

In addition to the direct characterization methods introduced above, we performed simulations to assist in identifying the key model features that reproduce the experimental phenomenology. 
Our simulations are not meant to mimic the experiment as accurately as possible, but to provide evidence regarding the primary mechanism with a model as simple as possible. 
Two questions are guiding our analysis: 
\begin{itemize}
    \item[1)] Do aggregates align under shear flow?
    \item[2)] Can weak preferential alignment cause anisotropy in the conductivity?
\end{itemize}

To tackle the first question, we adopt the same baseline model as introduced in Ref. \cite{Coupette-Percolation}, i.e., we consider stiff CB aggregates as the fundamental building blocks of our simulation. 
These aggregates consist of primary spheres of diameter $\sigma$ that are rigidly fused together in the course of a diffusion limited aggregation (DLA) process. 
As a consequence, aggregates are Brownian trees exhibiting a power-law decaying density profile which corresponds to a fractal dimension of $d_{\mathrm{f}} \approx 2.5$, similar to experimentally observed fractal dimensions of certain CB varieties (cf. \cite{kluppel1995fractal}). 
The tenuous agglomerate structures (secondary aggregates) are expected to get dispersed in the coating process and may or may not reform in a cured composite. 
However, as the samples are cured right away, we expect the resulting configuration of CB to resemble a steady state configuration of aggregates in shear flow.  
For simplicity, we assume that the primary function of the polymer matrix is mediating the momentum transfer between individual carbon black aggregates. 
Thus, we neglect the microscopic structure of the polymer in favor of including hydrodynamic interactions between carbon black aggregates. 

The DLA aggregates, though isotropically generated, are far from perfectly spherical. The gyration tensor $S_{nm}$ of an aggregate comprising $N$ primary particles
\begin{align}
S_{nm} = \frac{1}{N} \sum_{i = 1}^N \left(r_m^{(i)} - r^{\mathrm{CM}}_m\right)\left(r_n^{(i)} - r^{\mathrm{CM}}_n\right) \; ,
\end{align}
with $r_m^{(i)}$ denoting the $m^{\mathrm{th}}$ component of the position of the $i^{\mathrm{th}}$  primary particle and the center of mass position $\boldsymbol{r}^{\mathrm{CM}}$, can be diagonalized yielding, in analogy to the tensor of inertia, the principal axes of the aggregate. The corresponding principal moments $\lambda_i^2$  (eigenvalues of $S_{nm}$), $\lambda_1 \geq \lambda_2 \geq \lambda_3$, quantify the length of the principal axis of an ellipsoid with the same properties. This can be used to define an effective aspect ratio $\xi$ of an aggregate as $\xi = \sqrt{\lambda_1 / \lambda_3} \;$.

The aspect ratio provides a simple quantity to estimate how strongly the distribution of primary particles deviates from spherical symmetry. 
As more primary particles are added to an aggregate the influence of fluctuations in the growth process diminishes in comparison to the aggregate size and the average aspect ratio decreases with $N$. 
For our proof-of-principle simulations, we choose a uniform $N=20$ for which aggregates have an average aspect ratio of roughly 3. 
In reality the size of CB aggregates will be broadly distributed with a larger average aggregate size. 
However, smaller aggregates are easier to simulate and we expect any effects linked to the microscopic anisotropy of aggregates to be relatively more pronounced for smaller aggregates. 

Each aggregate has three distinguished axes that we label ``long'', ``middle'', and ``short'' axis, respectively. 
Likewise, the shear flow admits three distinguished directions, i.e., flow, gradient, and vorticity direction, respectively, just like in the experimental setup illustrated in Fig.~\ref{fig:tensile & 4PP specimens, 4PP polarity, planes+directions}. 
In order to quantify whether the aggregate axes adopt a preferential orientation under shear, we can define the nematic order parameter
\begin{align}
    S = \left\langle \frac{3 \cos^2(\theta) - 1}{2} \right\rangle \; ,
\end{align}
with $\theta$ denoting the angle between an axis of an aggregate and a direction of the shear flow. 
As a more nuanced indicator of anisotropy, we also depict the relative frequencies of the orientation of each axis as a histogram using a Mollweide projection.

\subsubsection{Aggregate alignment}

We simulate ensembles of aggregates in shear flow with Lees-Edwards boundary conditions using molecular dynamics (MD) with Multi-particle collision (MPC) dynamics \cite{malevanets1999mesoscopic, gompper2009multi, howard2019modeling}. MPC dynamics is a particle-based Navier Stokes solver, providing hydrodynamic interactions and thermal fluctuations using a mesoscale solvent that can be coupled to standard MD simulations. Details on this technique and the parameters we use for our simulations can be found in Appendix \ref{Appendix: Simulation}. 
As MPC simulations require substantial computational effort, these simulations are performed with only 125 CB aggregates, each consisting of 20 primary particles.

\begin{figure}
    \centering
    \includegraphics[]{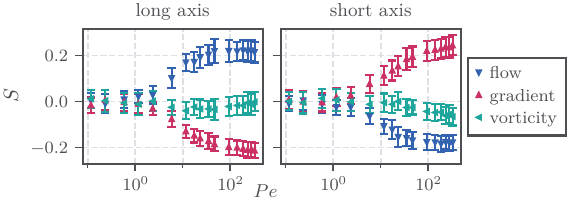}
    \caption{\justifying{Nematic order parameters of the long and short axes of aggregates with respect to the flow, gradient and vorticity axes for increasing P\'eclet number. Error bars encode the standard deviation.}}
    \label{fig:nop_vs_peclet}
\end{figure}

To quantify the influence of the shear flow, we introduce the P\'eclet number as $\Pe = {\dot\gamma R_g^2}/{D}$, with the shear rate $\dot\gamma$, the radius of gyration $R_g$ and the center of mass diffusion coefficient $D$ of the aggregates. 
$D$ can be measured via the aggregate mean squared displacement in equilibrium simulations. 
To assess whether CB aggregates align under shear, we conduct MPC-MD simulations for a broad range of P\'eclet numbers 
The alignment is measured by calculating the nematic order parameters with respect to the flow, velocity gradient, and vorticity axes, respectively, for increasing Péclet numbers. 
The results are illustrated in \autoref{fig:nop_vs_peclet}. 
ratio
For $\Pe=2.3$, where shear and diffusion have similar influence, and at lower shear rates, there is no discernible signature of alignment. 
However, starting from $\Pe=5.8$, the particles tend to align their long axis in the direction of flow. 
The negative order parameter in the gradient direction signifies that the long axes of the aggregates are preferentially orthogonal to the gradient. 
In combination with the vanishing nematic order parameter in vorticity direction, the overall picture indicates that aggregates tend to orient themselves in the vorticity-flow plane. 
This is supported by the nematic order parameters of the short axis, which frequently aligns parallel to the gradient and perpendicular to the flow direction.

To improve the visualization of this alignment, we use Mollweide projections of the aggregate axes in spherical coordinates. The polar angle $\theta$ corresponds to the gradient direction and is chosen such that the poles are represented by $\theta = \pm 90^\circ$. The azimuthal angle $\phi$ is measured with respect to the flow-gradient plane and thus encodes the vorticity direction at $(\phi, \theta) = (\pm 90^\circ, 0^\circ)$. 
The size of bins depends on the polar angle, especially noticeable near the poles of the Mollweide projection. To correct for this effect, we reweight the bin occupancies with $1 / \cos(\theta)$, the radius of circles of latitude.
\begin{figure}
	\centering
	\begin{subfigure}{.75\textwidth}
		\centering
		\includegraphics[width=\textwidth]{"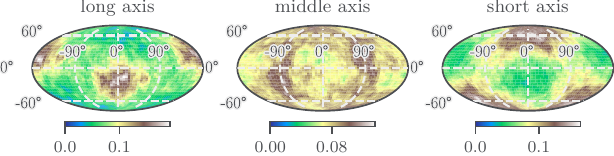"}
	\end{subfigure}
	\begin{subfigure}{.75\textwidth}
		\centering
		\includegraphics[width=\textwidth]{"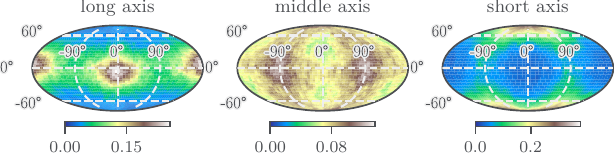"}
	\end{subfigure}
	\caption{\justifying{Mollweide projections of the distributions of the orientation of aggregate axes for weak ($\Pe =  5.8$, top) and strong flow ($\Pe = 93.4$, bottom), respectively. }}
	\label{fig:axes}
\end{figure}

Figure \ref{fig:axes} displays the distributions of the aggregate axes. It corroborates that the long axes of aggregates preferentially point in the flow direction and lie in the flow-vorticity plane. 
The short axis is parallel to the gradient direction, i.e., at the poles of the projection. The behavior of the middle axis is less clear, yet a maximum can be identified that is related to a slight alignment with the vorticity axis. 
The features of the distributions become more pronounced with increasing shear rate. There is also evidence of a symmetry-breaking effect affecting the polar angle. 
For example, the maxima of the distribution of the long axis are shifted away from the $\theta=0^\circ$ circle of latitude. 
This indicates long axes at a slight angle with respect to the flow-vorticity plane towards the gradient direction, as known experimentally \cite{vermant2001rheooptical} and theoretically \cite{winkler2004rod} for rods, an effect that is reduced for stronger flows.

Thus, we can conclude that aggregates show alignment in shear flow, exhibiting behavior similar to the phenomenology previously observed for aggregates fracturing in shear flow \cite{asylbekov2021microscale} as well as for  rod-like particles \cite{white2009simulations,rahatekar2005mesoscale,shi2014network,finner2018continuum}. Yet, the alignment effect, even in strong flow, is rather weak with nematic order parameters plateauing around 0.2.  
The microscopic randomness of the aggregates and their three distinct axes of symmetry do not appear to have a strong effect. 

\subsubsection{Anisotropic conductivity}

A macroscopically conductive CB composite relies on a percolating network of CB aggregates.
We can generate these networks from our simulations by assigning a threshold distance $d$.
If the surface separation between any two primary particles belonging to different aggregates undershoots $d$, we consider the respective aggregates connected.    
As we want to calculate the conductivity of the aggregate network, the length $d$ effectively represents a cutoff distance beyond which tunneling transport may be neglected. 
With this, we can interpret a system configuration as a random resistor network with aggregates as nodes and tunneling junctions as edges.
Introducing two electrodes connected to two opposite sides of the simulation box, we compute the resistivity of the composite by solving the Kirchhoff's equations.
The Kirchhoff network conductivity provides a crude indicator for the conductivity of the system as we assume that the intrinsic resistance of any aggregate is small compared to the resistance of tunneling contacts and any influence of surface chemistry or the polymer matrix is neglected.
Nevertheless, the Kirchhoff conductivity provides an indication of how the geometric arrangement of carbon black within the polymer matrix impacts the properties of the composite. 
Choosing the location of the electrodes, we can measure the conductivity of the network along different directions and compute the ratio $R_x / R_y$ of eq.~\eqref{eq:ratio}.
Thus, we can test whether the structural changes due to shear during coating induce anisotropy to the conductivity of the network. 

The MPC simulations are computationally demanding so that the system sizes that we can comfortably simulate are small. 
As a consequence, the conductivity measurements in the resulting networks are subject to strong fluctuations, obstructing a confident assessment of the electrical anisotropy.
Therefore, we perform computationally cheaper Monte Carlo simulations without hydrodynamic interactions but instead with an external alignment potential
\begin{align}
    V_A(\boldsymbol{\omega})  =  - A \boldsymbol{\omega} \boldsymbol{e}_z \; ,   
\end{align}
with the orientation of the long axis of an aggregate denoted as $\omega$ and the field strength $A$. 
These simplified simulations do not accurately capture the CB distribution in shear flow, in particular, because we only adapt the orientation of the long axis.  However, they still grant insight into the relationship between microscopic alignment of aggregates and conductivity.
As we have access to much larger system sizes and a higher number of independent snapshots to work with, our observations bear higher statistical significance.
%Subjecting the simulated aggregates to an affine transformation we may can also get a qualitative idea of how the conductivity changes in response to small strain.

We consider an ensemble of 1000 DLA aggregates comprising 20 primary particles each. Interaction between different aggregates is hard, i.e., no overlap is permitted between any two aggregates. As there is no simple way to set up aggregates without overlap, we equilibrate by initially allowing overlap at the cost of an energy penalty depending on the depth of interpenetration. We proceed by slowly lowering the temperature while also scaling the alignment potential so that it remains unaltered relative to the thermal energy. Once all overlaps are removed, the annealing potential is removed, followed by a second equilibration at unit temperature. As aggregates are rigid, they are characterized by three positional and three orientational degrees of freedom which are sampled through random translations and orientations of the entire aggregate. 
The system evolves in periodic boundary conditions but no conductive transport is allowed through the two walls chosen as electrodes.
For each alignment field strength $A$ and each density, we generate 100 independent configurations and analyse the corresponding Kirchhoff networks for resistivity parallel and perpendicular to the axis promoted by the alignment field. 
A more detailed description of the conductivity calculation and the chosen parameters can be found in appendix \ref{Appendix: Simulation}.
\autoref{fig:con_plot} illustrates the results. 

\begin{figure}
    \centering
    \includegraphics[]{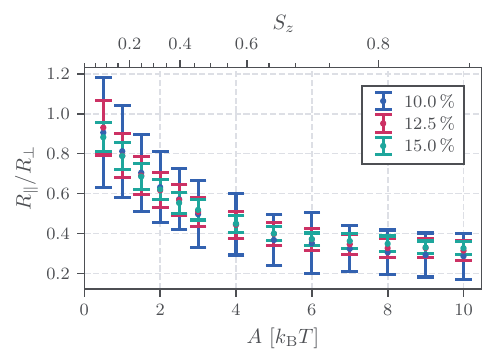}
    \caption{\justifying{Electrical anisotropy ratio induced by an external alignment field with amplitude $A$ for different filler densities. Points mark the mean ratio averaged over 100 networks and error bars show the corresponding standard variation. The nematic order parameter $S_z$ is a monotonic function of $A$ (unaltered across the three densities) and is depicted as the non-linear top axis. }}
    \label{fig:con_plot}
\end{figure}

For all field strengths, the average electric anisotropy ratio is smaller than one, indicating that conductivity is robustly enhanced in the direction of preferential alignment. The effect becomes more pronounced as the field strength is increased, inducing stronger alignment as measured by the nematic order parameter $S_z$ of the long axis relative to the alignment direction. An amplitude of $A=2\,k_\mathrm{B}T$ roughly reproduces the $S_z$ order parameter of the MPC-MD simulations for large P\'eclet numbers.
The absolute resistivities are subject to strong fluctuations and so is the ratio $R_\parallel / R_\perp$, with individual networks exhibiting ratios larger than one. These fluctuations diminish with $A$ as well as the density. As $10\,\text{vol-}\%$ is just above the percolation threshold without external alignment, the networks, particularly for weak alignment, have little redundancies and thus heavily rely on specific links that break and reform inducing fluctuations. These fluctuations primarily originate from the finite size of the simulation.
At larger densities, networks have a lot more connections, heavily reducing the impact of microscopic fluctuations.

The trend observed in simulations is again similar to previous studies on rod-like particles. However, it does not agree with our experimental observations that report an electric anisotropy ratio larger than 1 for unstrained films.

A more scrupulous glance at individual networks allows us to attribute the drop in the anisotropy ratio to the number and average length of paths linking the electrodes. Especially, at high field strengths the number of independent paths linking the parallel plates in alignment direction is substantially larger than for the perpendicular direction. Likewise, the average length of these independent paths is drastically reduced in alignment direction. A simple Fermi calculation based on the aforementioned metrics can adequately reproduce the Kirchhoff results. Thus, the network nuances like edge weights are largely inconsequential -- the network structure is sufficient to predict the trend.

Thus, whatever mechanism causes the experimental ratio to behave fundamentally differently, it exceeds the implications of a simple uniaxial aggregate alignment.%, consistent with the segmentation analysis.  

\subsection{Structural model for the observed (piezo-)electric anisotropy} \label{structural hypothesis}

Based on our results %from electrical, piezoresistive and structural characterization as well as simulations of CB aggregate alignment
, we hypothesize the following mechanisms to be responsible for the (piezo-) electric anisotropy of CB-silicone films generated by doctor blade coating:

In a first step, shear flow during doctor blade coating of the yet liquid CB-silicone mixtures fragments the CB network into (mobile) agglomerates and primary aggregates. As supportive evidence of this basic requirement for structural rearrangements, we know liquid precursors to be strongly shear-thinning in the concerned range of shear rates (see Section \ref{Local electrical properties (unstrained state)}). In addition, we have shown that in cured films, CB is dispersed uniformly within the silicone matrix (Section \ref{Nanomechanical mapping (PeakForce QNM)_results}), i.e., flow of matrix chain segments and of CB particles during coating is linked.

Subsequently, shear flow during coating has a two-fold effect:

a) CB aggregates are increasingly separated along the extensional axis of shear flow and, concomitantly, in the coating direction. This is supported by SAXS data for the silicone matrix which indicates more pronounced flow of chain segments parallel to the coating direction (mesh size in the coating direction is larger than perpendicular to it, Section \ref{SAXS_results}). Analogous evidence for CB aggregates is lacking so far due to the size limitation of the SAXS measurements, but CB aggregates should experience extensional flow similar to the silicone matrix due to their mechanical coupling. For this mechanism, electrical conductivity is expected to be lower in the coating direction than perpendicular to it, due to larger overall tunneling distances ($R_\parallel$/$R_\perp>1$).

b) Due to their anisometric nature (see Sections \ref{Nanomechanical mapping (PeakForce QNM)_results} and \ref{Segmentation and statistical analysis of nanomechanical data_results}, CB aggregates are preferentially aligned in the coating direction. This picture is only hinted at for near-surface CB aggregates and agglomerates (segmental analysis of nanomechanical data, Section \ref{Segmentation and statistical analysis of nanomechanical data_results}) but fully supported by simulations of CB aggregate alignment (Section \ref{Simulations_results}). The latter show that in the unstrained state, such alignment correlates with increased conductivity in the coating direction ($R_\parallel$/$R_\perp<1$). In addition, resistance should increase more strongly for strain perpendicular to the coating direction since the conductivity loss is more critical when preferentially aligned aggregates are pulled apart perpendicular to their major axis vs. parallel to it.

In the unstrained state, mechanism a) dominates such that electrical conductivity in the coating direction is lower than perpendicular to it, explaining the experimental anisotropy ratio of $R_\parallel$/$R_\perp>1$.
For strain-induced resistance changes, mechanism b) is the governing mechanism, accounting for the observed anisotropy in piezoresistive sensitivity, d$R_\perp(\epsilon_{\perp})/$d$\epsilon_{\perp}>$ d$R_\parallel(\epsilon_{\parallel})/$d$\epsilon_{\parallel}$. 

Concerning the dependence on shear rate, we assume structuration effects from both mechanisms to be close to maximal in our experiments since $R_\parallel$/$R_\perp>1$ in unstrained films did not notably depend on shear rate. Simulations support this picture of structuration effects becoming maximal at some point, as aggregate alignment and its resulting $R_\parallel$/$R_\perp$ saturate for large Peclet numbers.

\section{Summary, conclusions and outlook} \label{Summary, conclusions and outlook}

To our knowledge, this paper for the first time reports on flow-induced piezoresistive anisotropy of CB elastomers relevant for sensing applications, as well as underlying structure-property-relationships. The results help understand the impact of processing and CB concentration on final properties and thus provide a means to tune CB elastomers accordingly.

Doctor blade coating liquid CB-silicone mixtures with CB concentrations above the percolation threshold (7/9/11 vol\%) led to significant electrical anisotropy in cured films, with conductivity and piezoresistive sensitivity being higher perpendicular to the coating direction than parallel to it. %The coating process thus generates anisotropic CB structures which are maintained (largely or fully) during cure and storage.
Anisotropic electrical conductivity in unstrained films was seen for all examined compositions, gap heights and shear rates, stressing the practical relevance of flow-induced anisotropy for conductive CB-filled elastomeric films used in sensing applications. Even if electrical anisotropy appears weak in undeformed samples ($R_\parallel$/$R_\perp$ = 1.1 - 1.4 in our experiments, in tune with only weak structuration effects seen in simulations), it can strongly manifest upon stretching as a result of piezoresistive anisotropy (e.g. double as high for stretch parallel to the coating direction than perpendicular to it). Thus, as a practical conclusion,  process-induced (piezo-)electric anisotropy must be checked for, and processing and/or CB elastomer composition may need to be adjusted to minimize it. 
%Could however also use effect for sensing applications

Concerning structure-property-relationships, our results confirm trends for the impact of CB concentration on piezoresistivity and anisotropy found in literature on CB elastomers, and they can be motivated by fundamental considerations based on electrical percolation. In particular, the strain-induced resistance increase and its anisotropy (stretch parallel vs. perpendicular to the coating direction) both increased upon approaching the percolation threshold, consistent with the CB network becoming more sensitive to structural changes (less resilient) when only few conductive pathways are present. In contrast to the electrical response to strain, the mechanical stress response (along with matrix compressibility) was isotropic. We attribute this to the fact that electrical conductivity in CB-silicone films necessitates percolation, while the mechanical reinforcement of CB does not. In conclusion, electrical and mechanical effects of CB anisotropy in CB-filled elastomers can be very different in magnitude, and piezoelectric anisotropy can be minimized in practice by choosing CB concentrations further away (above) the percolation threshold.

%\begin{itemize}
  %  \item In undeformed films, electrical conductivity parallel to the coating direction is lower than perpendicular to it: vs. simplistic idea based on other anisometric fillers like carbon fibers, CNTs etc. being aligned in the coating direction / elongational axis of macroscopic shear. Electrical conductivity and the degree of its anisotropy in undeformed films were independent of global shear rate within a large interval (50 – 3000 s\textsuperscript{-1}). The mechanism leading to anisotropy is thus either not governed by dynamic effects, as long as shear forces are strong enough to break up the CB network. // Verbindung zu Simulation (Sättigung von Aggregatausrichtung für hohe Pe / Scherraten)?
%\end{itemize}

As a structural explanation of the observed (piezo-)electric anisotropy, we have derived the following hypothesis with the help of structural analysis of the silicone matrix and CB at the aggregate level (SAXS on bulk samples, nanomechanical mapping and segmental analysis for near-surface CB particles) as well as simulations on CB aggregate alignment:  Shear flow during coating fragments the CB network and then has a two-fold effect, i.e., it
induces a) preferential aggregate alignment, as well as b) increased interparticle distances, parallel to
the coating direction. When unstrained, mechanism b) dominates such that $R_\parallel/R_\perp>1$. In contrast mechanism a) dominates upon straining such that the resistance increase is stronger for stretch perpendicular to the coating direction (and the direction of preferential aggregate alignment, respectively). 

%o	silicone matrix weakly anisotropic when unstrained (SAXS), with no measurable mechanical anisotropy analogous to macroscopic mechanical isotropy

%o	PeakForce QNM clearly maps dispersion state of near-surface, with resolution down to aggregate level. Can (nanomechanically) explain reinforcement effect by CB

%o	segmentation of PeakForce QNM maps and statistical analysis: confirms that aggregates are not broken down by doctor blade coating

In conclusion, the proposed hypothesis is fully consistent with the observed phenomenology and with trends seen in simulations. Among others, it shows that the common picture on the impact of shear flow on anisometric particles, i.e., preferential alignment in the flow direction, is too simplistic.
Since evidence of anisotropy up to the aggregate level is only weak overall, future work should tackle the larger length scales up to the CB network level. For example, our hypothesis could be substantiated by structural analysis of the yet liquid, in situ-sheared material via rheo-USAXS (fresh CB-silicone mixtures in stationary shear flow at gaps of $10^2$ µm) as well as cured films via USAXS and FIB-SEM tomography (combined with network reconstruction and statistical analysis). 

%\begin{itemize}
 %   \item simulations
  %  \begin{itemize}
      %  \item Aggregates show weak alignment in shear flow with the long axis of an aggregates preferentially oriented in the flow-vorticity plane. 
     %   \item Alignment saturates for large Peclet numbers.
     %   \item Preferential alignment of the long axis induces anisotropy in the conductivity, however, with reduced conductivity perpendicular to the director.
      %  \item Anisotropy becomes more pronounced with stronger alignment.
  %  \end{itemize}
  %  \end{itemize}

\newpage
\section{Acknowledgments}

Funded by the Deutsche Forschungsgemeinschaft (DFG, German Research Foundation) under Germany's Excellence Strategy – EXC-2193/1 – 390951807.
Funded by the Deutsche Forschungsgemeinschaft (DFG, German Research Foundation) - 404913146, 457534544, 531007218.
The authors acknowledge support by the state of Baden-Württemberg through bwHPC
and the German Research Foundation (DFG) through grant no INST 39/963-1 FUGG (bwForCluster NEMO).
%We thank the DFG for grant 531007218.
%We thank the DFG for grant 457534544.
We thank Marisol Ripoll and Roland Winkler for helpful discussions regarding MPC.

Furthermore, we thank Werner Schneider and Herbert Beermann for manufacturing the four-point probe setup, Lola González-García and Dominik Schmidt for helpful discussions, and Prof. Christian Motz for providing his AFM for the PeakForce QNM measurements.
 
There are no conflicts of interest to declare.

\newpage
\appendix
\begin{appendices}
\label{Appendix/SI}
\renewcommand{\thesection}{\Alph{section}}
\renewcommand\thesubsection{\thesection\arabic{subsection}}
\counterwithin{figure}{section}
\section{Film fabrication (detailed description)} \label{Film fabrication (detailed description)}

 CB by ThermoFisher Scientific (Carbon black, acetylene, 100 \% compressed, 99.9+\%, Thermo Scientific Chemicals) was dispersed in a silicone resin (base component of Sylgard\textsuperscript{®} 184 silicone elastomer, Dow\textsuperscript{®}) via speedmixing (SpeedMixerTM DAC 600.2 VAC-P by Hauschild, 3 min at 2350 rpm under vacuum). The silicone crosslinker (curing agent of Sylgard\textsuperscript{®} 184) was then added with the concentration recommended by the manufacturer (1/10 of the mass of the base component \cite{Sylgard-TDS}). Final dispersion was achieved by subjecting the mixture to the aforementioned speedmixing protocol. Three compositions above the percolation threshold ($\sim$5 vol\% CB \cite{Coupette-Percolation}) were realized: 7/9/11 vol\% CB. Volume percentages refer to the volume of the resin and were calculated using the densities reported in [22] ($\rho_{CB}$ = 1.7 g/cm$^3$, $\rho_{resin}$ = 0.965 g/cm$^3$).

\begin{figure}
    \centering
    \includegraphics[width=0.85\linewidth]{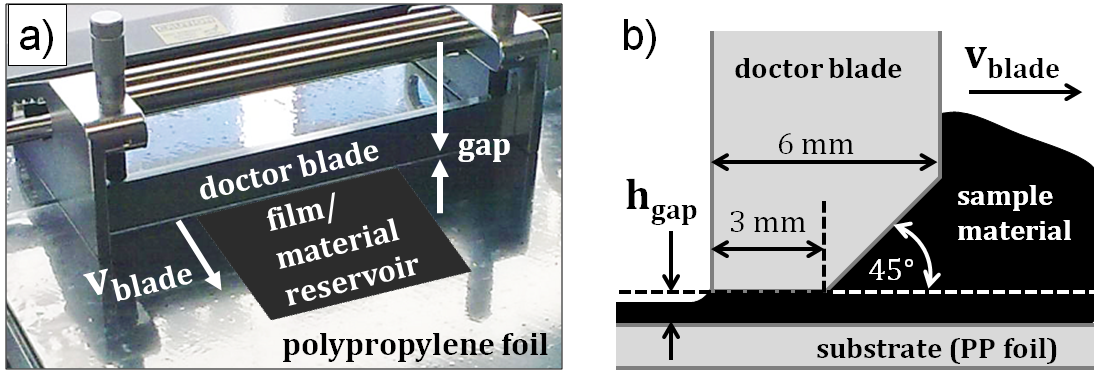}
    \caption{Doctor blade coating: a) overall setup b) coating geometry}
    \label{fig:doctor blading apparatus}
\end{figure}

Doctor blade coating was done in ambient air (22 °C, 39 ± 13 \% r.h.) immediately after mixing, using a film applicator (automatic film applicator by TQC) with the doctor blade setup (Microm II Metric Film Applicator by Gardco ®) depicted in Fig. \ref{fig:doctor blading apparatus}. The reactive mixture was placed in front of the doctor blade with a spatula such that mechanical manipulation is minimal and the sample amount suffices to achieve lateral film dimensions of at least 5 x 7 cm$^2$. Two gap heights, $h_{gap}$ = 60 µm and 350 µm, were realized by manually adjusting the micrometer screw of the coating apparatus. In order to assess the impact of dynamic effects during processing on final properties, the blade speed, $v_{blade}$, was varied between 5 mm/s and 400 mm/s. The resulting global shear rates ($\gamma_{global}$= $v_{blade}$/$h_{gap}$) span a large interval of 50 s$^{-1}$ to 3000 s$^{-1}$. Polypropylene foil was used as a substrate because of its insulating properties (for electrical four-point probe measurements) and easy detachment of films (relevant for the ‘thick’ films used for tensile tests).

Curing took place in a convection oven for 2~h at 100~°C. Films were prepared using the desired process conditions and removed from the film applicator. Subsequently, the doctor blade was cleaned and the next film was prepared. All films were transferred to the oven as soon as the last film was coated. The resulting time between coating and curing ranged between 10 min (last film) and 45 min (first film). The optical appearance of the films changed neither during this waiting time nor during the curing step, indicating that macroscopic flow after coating was marginal.

The mean thickness of cured films ($d_{film}$, determined by light microscopy of cross-sectional points) was in the range of 80 – 100~µm for the 60~µm gap and of 260 – 350~µm (see next paragraph for respective ranges for 7/9/11~vol\%~CB ) for the 350~µm gap. The thickness of the 60 µm-films is thus systematically higher than the gap height, while the thickness of the 350~µm-films tends to be lower. The former can be traced back to inaccuracies in the adjustment of the micrometer screws (offset at the zero-position); the latter is likely due to the films thinning out at the blade exit (as indicated in Fig. \ref{fig:doctor blading apparatus}b, typical of shear-driven flow  \cite{Kim-Analytical}) as well as polymerization shrinkage.

In addition to the conductive CB-silicone films, a neat silicone film (0 vol\% CB) was fabricated to give specimens for mechanical (uniaxial tensile test) and structural (SAXS) characterization of the silicone matrix. Just as in the case of CB-silicone specimens for electromechanical and structural analysis, the blade speed was set to 20~mm/s. Due to the much lower viscosity compared to CB-containing mixtures, coating did not take place immediately after mixing resin and hardener, but the mixture was allowed to polymerize till viscosity was similar to the 7 vol\% CB-mixture (4~h in ambient air + 5~min at 100°C + cooling down for some minutes). Together with a slightly higher gap height of 400~µm (vs. 350~µm), the resulting film was comparable in thickness (260 - 280 µm) to CB-silicone films coated at 350~µm, 20~mm/s (260 - 290 µm / 280 - 320 µm / 320 - 350 µm for 7/9/11~vol\%~CB).

\section{Processing of electromechanical raw data} \label{appendix: electromechanical}

\begin{figure}
    \centering
    \includegraphics[trim={0cm 0.78cm 0cm 0cm},width=0.532\linewidth]{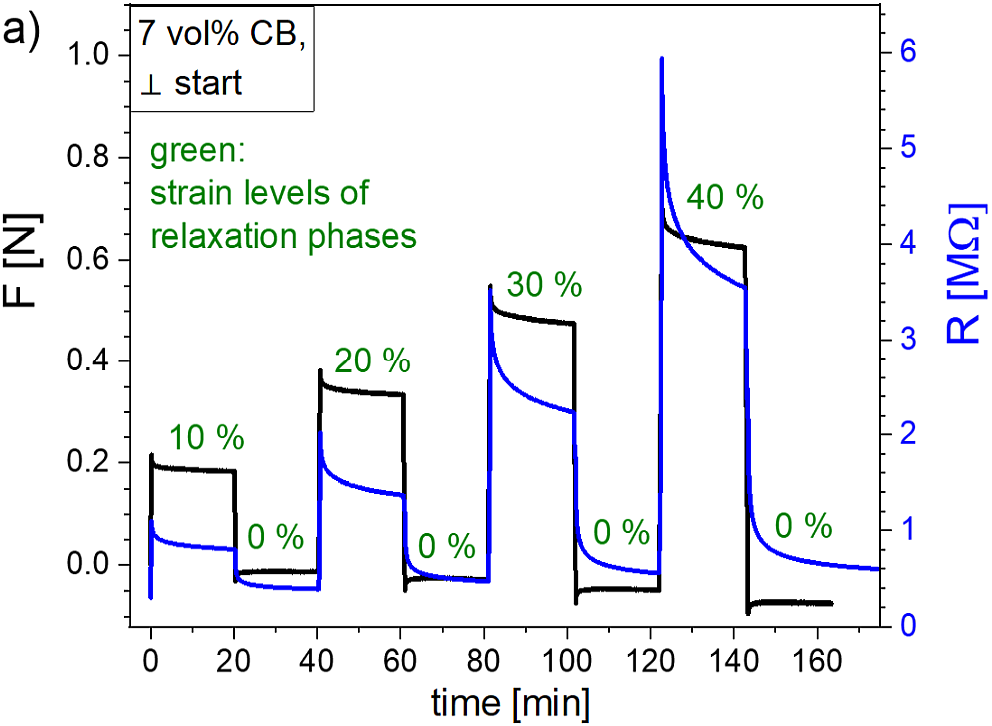}
     \includegraphics[trim={0cm 0.78cm 0cm 0cm},width=0.549\linewidth]{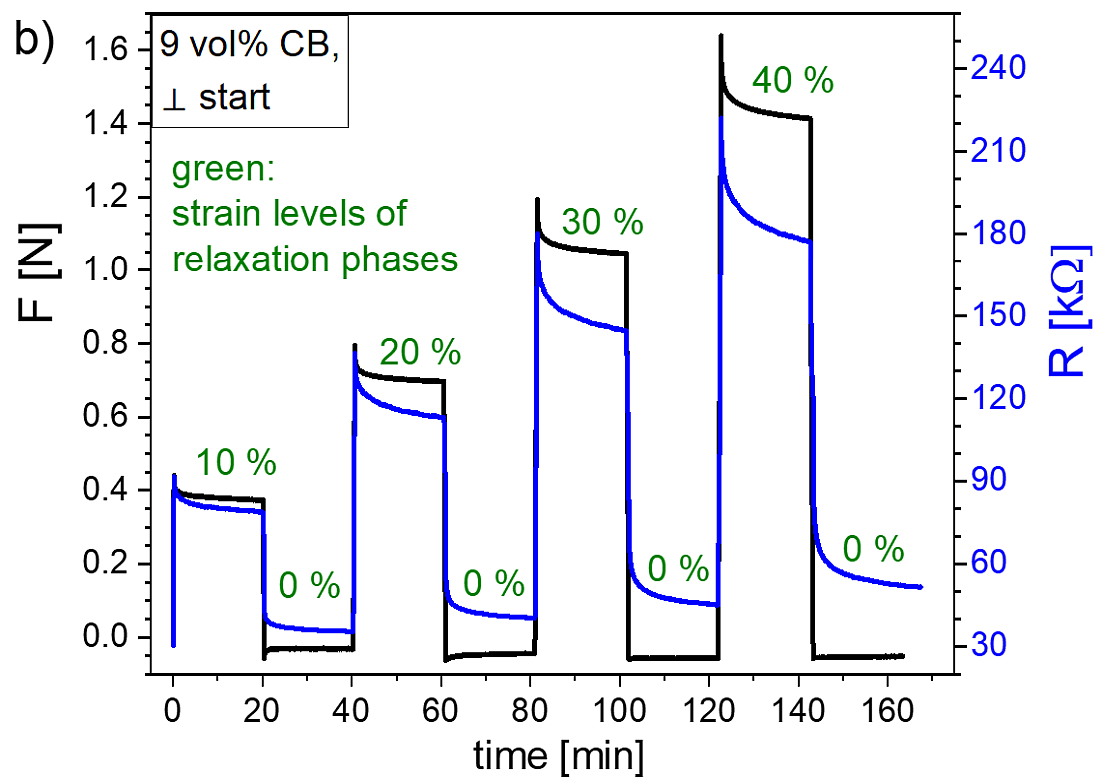}
    \includegraphics[width=0.536\linewidth]{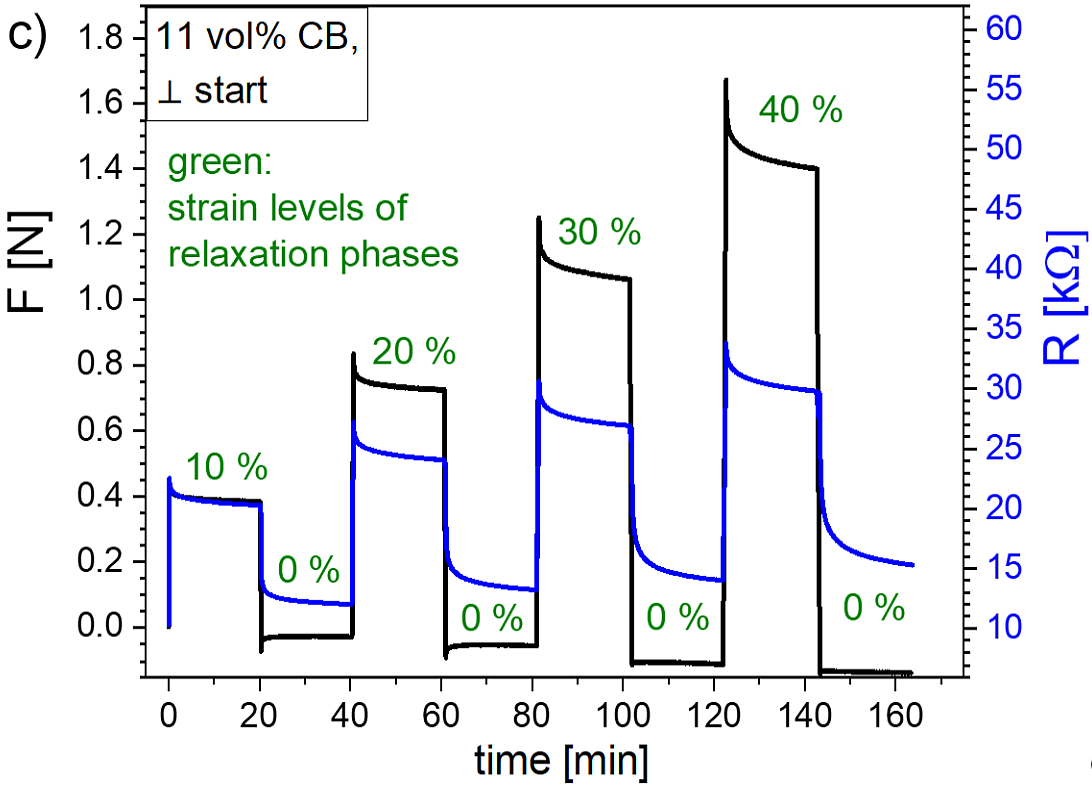}
    
    \caption{\justifying{Examples of raw data of uniaxial tensile tests (load-unload cycles to  strain plateaus $\epsilon$\textsubscript{max} = 0/10/20/30/40 \%, 10\textsuperscript{-2} s\textsuperscript{-1} strain rate, 20 min hold time at each strain level, 22°C ± 1~K, 30~±~15~\% r.h.) with in-situ electrical two-point measurement at room temperature: electrical resistance, R, and tensile force, F, for stretch perpendicular to the coating direction of samples with 7/9/11 vol\% CB. Curves of samples stretched parallel to the coating direction bear the same qualitative features.}}
    \label{fig:force & resistance raw data}
\end{figure}

%anpassen + Graphen aller relevanten Rohdaten dazu
The electrical resistance response to uniaxial strain at room temperature, just like the mechanical stress response, involves pronounced relaxation (see portions at the strain plateaus in Fig. \ref{fig:force & resistance raw data}). For the first strain plateau (10~\%), the hold time (20~min) suffices for tensile force and electrical resistance to fully relax (constant values after loading to 10~\% as well as after unloading back to 0~\%). For higher strains, force and resistance plateaus are not yet reached. To approximate relaxed resistances, the relaxation phases were fit in Origin 2022 using eq. \eqref{eq:exponential fitting}. The desired plateau value equals $R_\mathrm{relax}$.
\begin{equation} \label{eq:exponential fitting}
R=R_\mathrm{relax}+A_1 \exp(-(t-t_0)/\tau_1 )+A_2 \exp(-(t-t_0)/\tau_2 ) +A_3 \exp(-(t-t_0)/\tau_3 )	
\end{equation}

\begin{figure}
    \centering
    \includegraphics[width=0.5\linewidth]{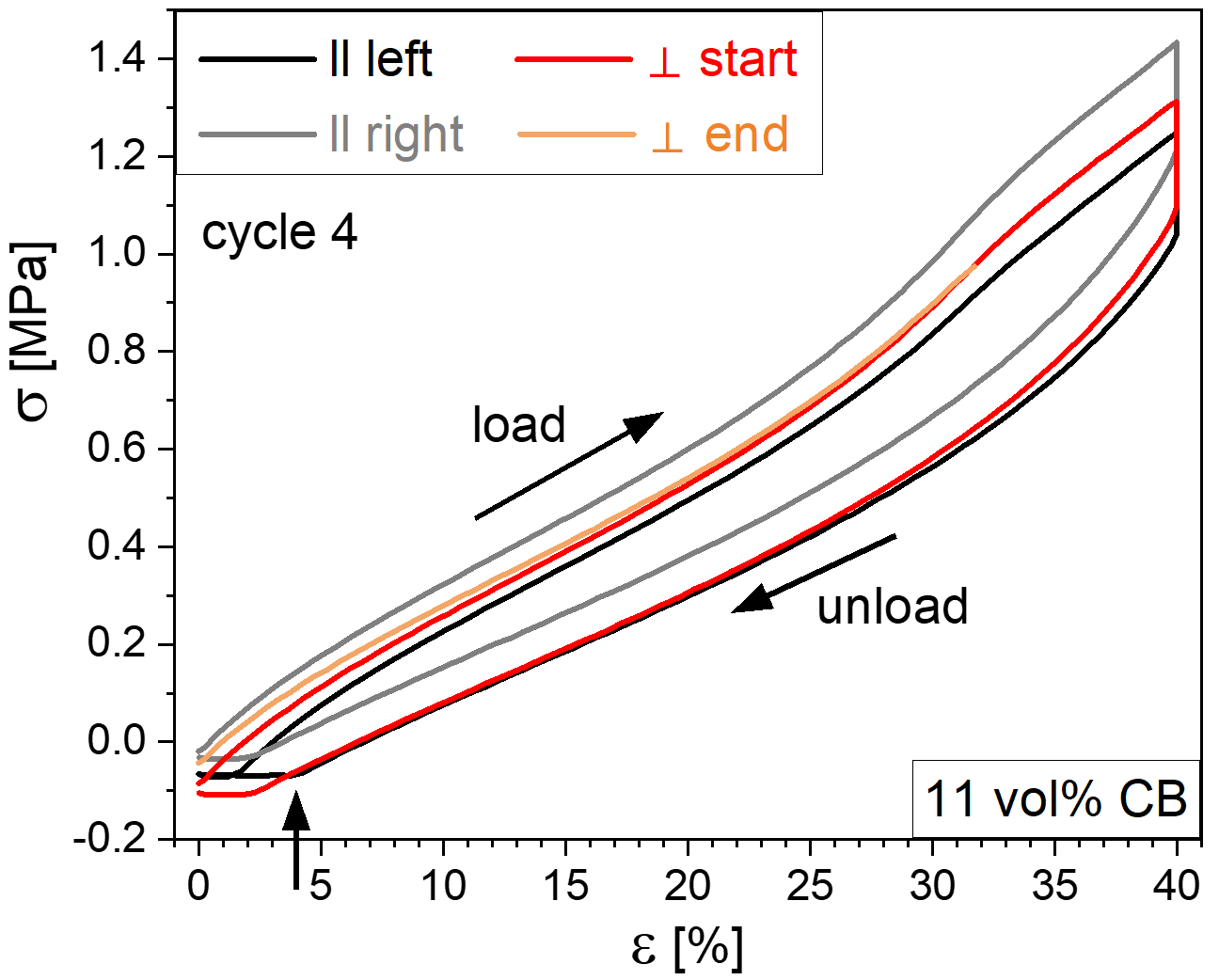}
    \caption{\justifying{Unshifted stress-strain curves of uniaxial tensile tests (22 °C ± 1 K, 30 ± 15 \% r.h.), cycle 4 (load to \(\epsilon\) = 40 \%, relax, unload), for 11~vol\% CB and stretch parallel or perpendicular to the coating direction, respectively. The expressions 'left', 'right', 'start' and 'end' refer to the positions on the film from which the tensile specimens were cut (see Fig. \ref{fig:tensile & 4PP specimens, 4PP polarity, planes+directions} in Section \ref{Characterization}). The vertical arrow indicates the maximal plastic strain seen for all samples (from specimen cut from left side of the film, stretched parallel to the coating direction).}}
    \label{fig:stress-strain curves unshifted}
\end{figure}

From the force curves in Fig. \ref{fig:force & resistance raw data} it can be discerned that the tensile force generally drops below zero after unloading to 0 \% strain, and stays negative after stabilization. This effect is related to plastic deformation and/or slip which reduce the pre-load. For optimal comparison of the stress response, the stress-strain curves presented in the following have been shifted to zero force and zero strain at the beginning of each load phase. According to the unshifted curves, plastic deformation and/or slip becomes apparent only for CB-filled samples (not the neat silicone) and only after loading to 30 \% strain (not yet for the 10/20 \% strain plateaus). As shown in Fig. \ref{fig:stress-strain curves unshifted} for the ‘extreme’ case (11 vol\% CB, 4\textsuperscript{th} cycle to 40 \% strain), plastic deformation and/or slip cause a permanent nominal strain of \(\epsilon\) = 4 \% at the most. 
  
\section{Sensing performance} \label{Sensing performance}
In light of the applicability of CB elastomers as highly deformable resistive strain sensors, we assess our films in terms of linearity, sensitivity and durability/cyclability of the piezoresistive response. Their peculiarity would be their piezoresistive anisotropy, a property which could be exploited to create one-piece sensors with (bi-)directional sensitivity. 

\begin{figure}
    \centering
    \includegraphics[width=1\linewidth]{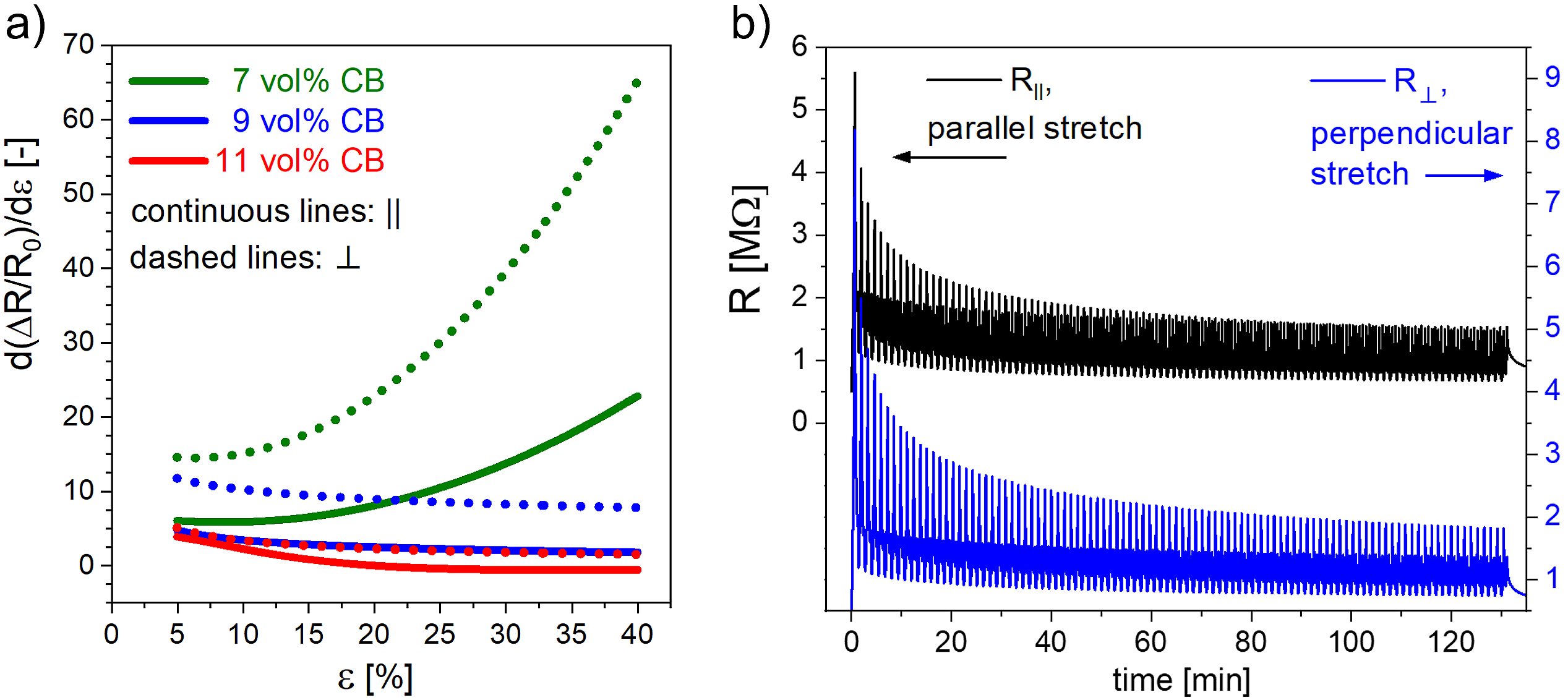}
    \caption{\justifying{To assess the CB-silicone films as a potential sensor material, piezoresistive sensitivity was quantified by differentating the trend lines of the reversible resistance increase in Fig. \ref{fig:piezoresistivity compact}b-f with respect to strain (equivalent to the gauge factor) parallel and perpendicular to the coating direction, respectively (panel a). To eliminate irreversible effects, one 7 vol\% CB-sample for each stretch direction was subjected to cyclic conditioning (at 22 °C ± 1 K, 30 ± 15 \% r.h., see main text for details) which results in resistance responses that become stationary toward the end of the procedure (panel b).}}
    \label{fig:gauge factor}
\end{figure}

To evaluate linearity and sensitivity, Fig. \ref{fig:gauge factor} shows the first derivative of each trend line for the reversible resistance increase in Fig. \ref{fig:piezoresistivity compact}b-e, d($\Delta R/R_0$)/d$\epsilon$, which is equivalent to the gauge factor for infinitesimally small strains ($\Delta R/R_0/\epsilon$). Since the values are not constant for any CB content and stretching direction in any meaningful interval, the films would perform non-linearly and would need to be calibrated accordingly. Piezoresistive sensitivity increases upon approaching percolation, with gauge factors as high as 20 – 25 (parallel) and 60 – 70 (perpendicular) for our lowest CB content of 7~vol\%. According to the review in \cite{Chen-PDMSbased}, these values compete with the majority of gauge factors reported for silicone-based piezoresistive sensors (mainly between 1.2 and 29, $>$ 100 in only a few exceptional cases involving graphene or carbon nanotubes). Thus, the films would give sensors with good sensitivity and minimal expenses for the filler (low-cost commercial CB, vs. CNTs, silver nanoparticles etc.).

Concerning the cyclability of the piezoresistive response, the irreversible resistance increase is too high in that it leads to significant differences between the overall resistance increase and the reversible part (except for perpendicular stretch of 7 vol\% samples, compare corresponding curves in Fig. \ref{fig:piezoresistivity compact}b-e, Section \ref{Electrical resistance response to uniaxial strain}). This calls for cyclic mechanical preconditioning, a common method to eliminate irreversible effects (see \cite{Flandin-Anomalous,Kost-Resistivity,Huang-Strain,Yamaguchi-Electrical} for examples of CB-silicone elastomers). As a first attempt, we conditioned two specimens of the most sensitive composition (7 vol\% CB, one sample for each stretching direction) by subjecting them to 100 load-unload cycles between 1~\% and 40~\% strain (same strain rate as before, 10$^{-2}$~s$^{-1}$). The corresponding resistance responses are shown in Fig. \ref{fig:gauge factor}b.

\begin{figure}
          \centering
   \includegraphics[width=1\linewidth]{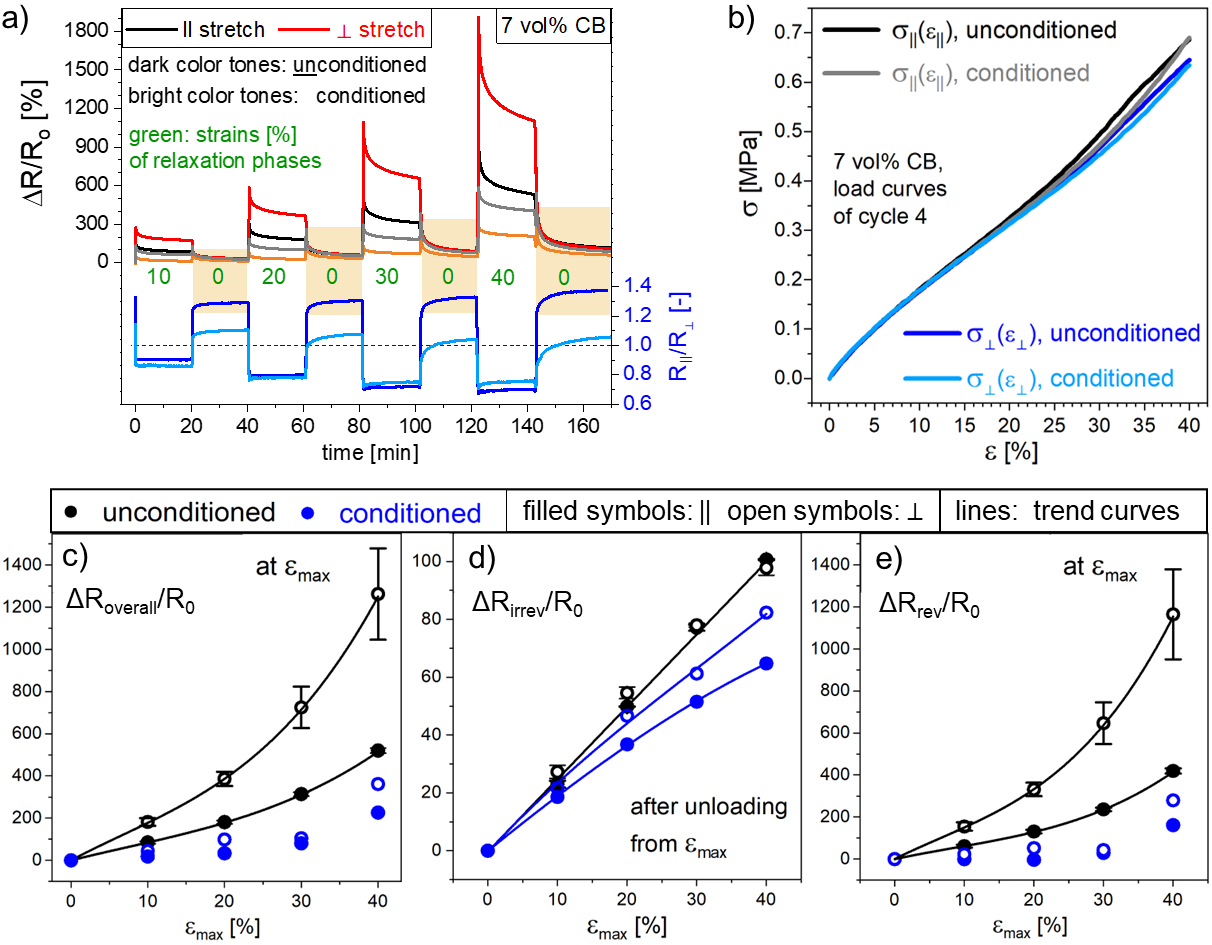}
          \caption{\justifying{Cycling conditioning hardly modified mechanical properties but clearly reduced both irreversible and reversible strain-induced resistance changes (relative to initial value at zero strain, $R \textsubscript{0}$) during subsequent tensile tests, as shown by the comparison of conditioned and unconditioned 7 vol\% CB-samples with regard to: a) electrical resistance change and anisotropy ratio during load-unload cycles (strain to $\epsilon \textsubscript{max}$ = 10/20/30/40 \%, hold at $\epsilon \textsubscript{max}$ for 20 min), b) stress response to strain of cycle 4 ($\epsilon \textsubscript{max}$ = 40~\%), c) overall resistance change to uniaxial strain (relaxed value at $\epsilon \textsubscript{max}$), d) irreversible resistance change (relaxed value after unloading from $\epsilon_{max}$), and e) reversible resistance change (overall change minus irreversible part). All measurements done at 22 °C ± 1 K, 30 ± 15 \% r.h.}}
          \label{fig:step-relaxation after cycling}
      \end{figure}

For both stretching orientations, resistance changes most during the first cycle and approaches a stationary regime toward the end of the procedure. Based on this, we could expect irreversible resistance changes to have ceased. To verify, we again conducted load-unload tensile tests to $\epsilon \textsubscript{max}$ = 10/20/30/40 \% with hold times of 20 min. The stress response and the relaxed values of the relative resistance change of the conditioned samples are displayed in Fig. \ref{fig:step-relaxation after cycling}, along with the data for the unconditioned ones. 
      
The stress response is barely altered by the cycling procedure, with stress-strain curves being identical up to $\epsilon$ $\approx$ 20~\%, and the stress having dropped only slightly (by $\leq$ 1/10) for higher strains (Fig. \ref{fig:step-relaxation after cycling}a). In contrast, the modification of piezoresistivity is pronounced (Fig. \ref{fig:step-relaxation after cycling}c-e). This again shows that electrical conductivity of the composite is much more sensitive to microstructural changes than its (visco-)elasticity, probably because the former relies on percolation while the latter does not (see Section \ref{Electrical resistance response to uniaxial strain}). The irreversible resistance increase is, as intended, reduced by cyclic conditioning (Fig. \ref{fig:step-relaxation after cycling}c); however, the reversible part of the strain-induced resistance increase drops even more in comparison (Fig. \ref{fig:step-relaxation after cycling}d). As a result, the irreversible part makes up a bigger portion of the overall resistance increase, and piezoresistive sensitivity (incl. its anisotropy) is greatly reduced overall. In conclusion, our films are not yet suitable as sensor materials, and more work on the stabilization of the piezoelectric response is needed. In addition, tests on the long-term stability of the films apart from mechanical conditioning are necessary to verify whether the flow-induced anisotropy prevails during relevant lifetimes, or whether it is significantly affected by material aging.

\section{PeakForce QNM maps and value distributions} \label{Appendix: PeakForce QNM maps and value distributions}
\subsection{General information on PeakForce QNM signals} \label{General information on PeakForce QNM signals}
\begin{figure}
    \centering
    \includegraphics[width=0.75\linewidth]{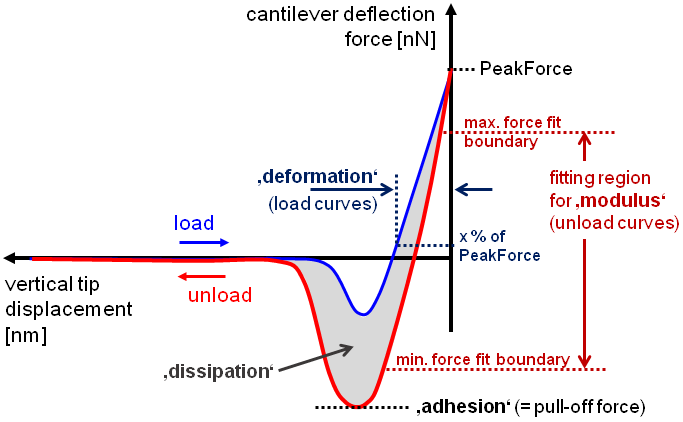}
    \caption{\justifying{Visualization of PeakForce QNM signals as derived from force-distance curves (deflection force of the cantilever as a function of vertical tip displacement) by the real-time analysis during the measurement. The reference point of the vertical tip displacement (zero point) is the lowest position, i.e., when indentation is maximal. In the absence of significant sample relaxation during loading, the PeakForce is reached at this point.}}
    \label{fig:PF-QNM force-distance curve signal evaluation}
\end{figure}

This section is intended to introduce the reader to the origin and informational value of the PeakForce QNM signals, named ‘deformation’, ‘adhesion’, ‘modulus’, and ‘dissipation’. Further information is found in publications of the patent holder Bruker \cite{Hua-PeakForce-QNM,Pittenger-Nanoscale,Pittenger-Quantitative}. We shall explain the signals using the schematic in Fig. 22. They are derived for each pixel (real-time analysis during scanning) from the load/unload curves that result from indenting the sample with the SFM tip. Contrary to the force-distance curves displayed by the measuring software, the abscissa of the analyzed curves is not the vertical position of the cantilever (z) but the vertical displacement of the tip. Only the latter is fully translated into the indentation depth ($\delta$, corresponding to tip displacement in the contact region, see Fig. \ref{fig:PF-QNM force-distance curve signal evaluation}) that is needed for evaluating the deformation and modulus signals. In contrast, the vertical displacement of the cantilever does not only lead to the indentation of the sample but also to the deflection of the cantilever, such that the z-values overestimate $\delta$ in the repulsive region and underestimate it in the attractive region.
 
The deformation signal is an approximation of the maximal indentation depth. The latter is equal to the vertical tip displacement between the first tip-sample contact (usually marked by a ‘snap-in’ of the tip into the sample as a result of attractive interactions) and the lowest tip position (at maximal indentation) of the load curve. The latter usually corresponds to the maximal repulsive force (= PeakForce), unless sample relaxation is so strong that the maximal force is already reached before maximal indentation. Since the contact point is not always readily discerned, the software identifies a point with a low repulsive force instead (e.g. 5 \% of the PeakForce), and outputs the vertical tip displacement between this point and the point of maximal indentation. Accordingly, deformation values  tend to be smaller than the true maximal indentation depth.

The adhesion signal is equal to the pull-off force, i.e., the maximal attractive force during unloading which must be overcome to sever the tip-sample contact. The term ‘adhesion’ is misleading since the pull-off force is not only the result of adhesive tip-sample interactions in equilibrium, but of dynamic effects, roughness, and additional attractive interactions (e.g. from liquid meniscuses and contaminations). One important methodological consequence is the sensitivity of the adhesion signal to humidity (when measuring in air) and contaminations. The latter usually increase the pull-off force and can lead to a poor reproducibility of absolute values.

The modulus signal is a mechanical stiffness value obtained by fitting the unload curve with the so-called DMT model. The latter accounts for adhesive tip-sample interactions in the form of a force offset equal to the pull-off force, and it is valid only for homogeneous, smooth elastic solids. Consequently, the modulus signal generally outputs effective stiffness values rather than the Young’s modulus. For polymer composites, the term ‘effective’ comprises the influence of both heterogeneity and viscoelasticity (dynamic contributions to the mechanical response of the sample). In the case of our CB-silicone composites, the modulus signal appears blurred (reduced lateral resolution) because it averages over heterogeneities with extreme local stiffness differences (i.e., regions where the stiff CB particles and the resilient silicone matrix both significantly contribute to the material response), and because the modulus signal is dominated by the contact region where the deformed sample volume is largest.

The dissipation signal is equal to the dissipated mechanical energy per oscillation period. It is obtained by evaluating the hysteresis area between the load and unload curves of a given load-unload cycle, i.e., by integrating their force difference over the traveled distance of the cantilever. It is usually the most robust of all signals because baseline fluctuations cancel out when integrating the difference of the unload curve and the load curve.

\begin{figure}
    \centering
    \includegraphics[width=0.95\linewidth]{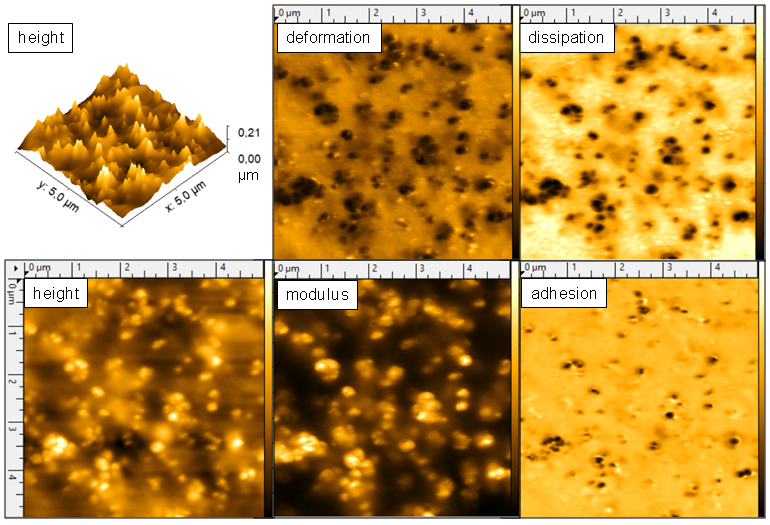}
    \caption{\justifying{PeakForce QNM maps mirror the dispersion state of near-surface CB, thanks to the material contrast between CB and the silicone matrix (in addition to topology). As an example, the maps of an unstretched 11 vol\% CB-film are shown (scan size 5x5 µm$^2$, coating direction from bottom left to top right, explanations on the signals see main text).}}
    \label{fig:PF-QNM_all signals overview}
\end{figure}

To explain PeakForce QNM signal maps and derive nanomechanical information on our CB-silicone films, Fig. \ref{fig:PF-QNM_all signals overview} gives an example of a 5x5 µm$^2$ scan on a film with 11 vol\% CB. We omitted the signal values to obtain a compact representation but assure that they are physically reasonable as well as reproducible (see Appendix \ref{Appendix: PeakForce QNM maps and value distributions} for a quantitative in-depth discussion). Keep in mind that PeakForce QNM yields near-surface information. In our case, the probed regions extend to some $10^1$ nm to $10^2$ nm into the material (see Section \ref{PeakForce QNM (quantitative nanomechanics)_experimental}).
 
The height signal (= sample topography) reveals a ‘hill-valley’ surface morphology. With the help of the other signals, the ‘hills’ are unambiguously attributed to regions rich in CB and the ‘valleys’ to regions dominated by the silicone matrix: The ‘hills’ are less deformable (lower values of the deformation signal) and stiffer (higher modulus values) than the surrounding valleys, as expected for embedded CB in light of its reinforcing effect (see Section \ref{Mechanical stress response to uniaxial strain}). The ‘valleys’ yield higher dissipation values because the silicone matrix reacts with pronounced viscoelasticity whereas regions dominated by CB do not (see Appendix \ref{Appendix: PeakForce QNM maps and value distributions} for further discussion). In contrast to deformation and modulus, the adhesion signal (= pull-off force) does not correspond to the ‘hill-valley’ morphology, as a large portion of the CB-rich regions is ‘invisible’. The reason for this lies in the pull-off force being governed by adhesive interactions between the tip and the sample surface rather than the mechanical behavior of the indented volume. Since CB particles are always silicone-coated in the initial reactive mixture and since their adhesion to the silicone matrix is good according to the mechanical reinforcement by CB, it stands to reason that most CB particles at the sample surface are covered by at least a thin layer of silicone. As a result, a big portion of CB-rich regions yields similar adhesion values as the silicone-rich regions.

In conclusion, PeakForce QNM allows a clear discrimination between the silicone matrix and embedded CB particles thanks to their different mechanical behavior as reflected by the deformation, dissipation and modulus signals. Concerning lateral resolution, the modulus signal is more blurred and thus less suited for structural analysis than the deformation and dissipation signals (see Fig. \ref{fig:PF-QNM_all signals overview}, plus Appendix \ref{Appendix: PeakForce QNM maps and value distributions} for explanation). We shall therefore focus on deformation and dissipation for discussing CB morphology in both unstretched and stretched states.

\subsection{Quantitative discussion of PeakForce QNM data} \label{Quantitative discussion of acquired PeakForce QNM data}
\begin{figure}
     \centering
     \includegraphics[width=0.7\linewidth]{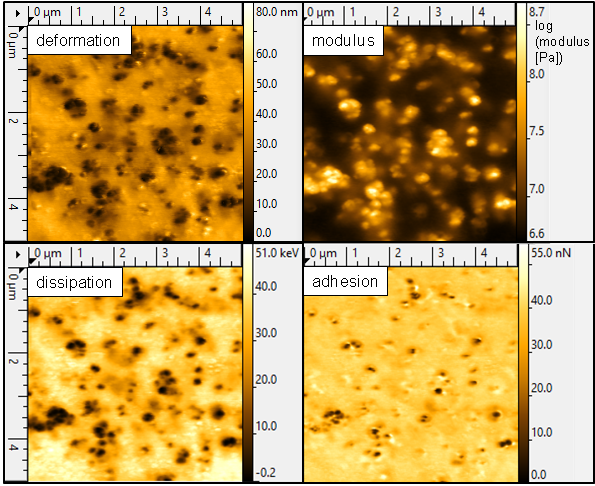}
     \caption{\justifying{Example of scaled maps of the PeakForce QNM signals ‘deformation’, ‘modulus’ (logarithmic scaling), ‘dissipation’, and ‘adhesion’ (11 vol\% CB, unstretched, 5x5 µm\textsuperscript{2} scan size)}}
     \label{fig:PF-QNM signal maps typical scaling}
 \end{figure}

In this section, we provide a quantitative discussion of PeakForce QNM signals along with further nanomechanical information on the CB-silicone films. For this we analyze respective value distributions (frequency densities) of repeatability measurements (scans of 10x10 µm$^2$) for various CB contents (7/9/11 vol\% CB) and material states (unstretched/stretched). To roughly assign values to the silicone matrix and CB particles, respectively, Fig. \ref{fig:PF-QNM signal maps typical scaling} gives an example of signal maps with typical value ranges. Compared to the silicone matrix, regions dominated by CB particles yield low deformation values of about 0~–~20 nm, low dissipation values of about 0~–~20~keV, low adhesion values of about $\leq$~20~nN, and high modulus values of some 10~–~100~MPa. (The color scale for the modulus is logarithmic in order to clearly reveal the filler-matrix morphology. In contrast, a linear color scale results in very dark images with a few bright spots, due to the huge stiffness differences between matrix and filler.) Values for the silicone matrix are more difficult to define since the transition between ‘filler-dominated’ and ‘matrix-dominated’ is very broad, due to the various degrees of spatial overlap within the deformed volume.

\begin{figure}
     \centering
     \includegraphics[width=0.45\linewidth]{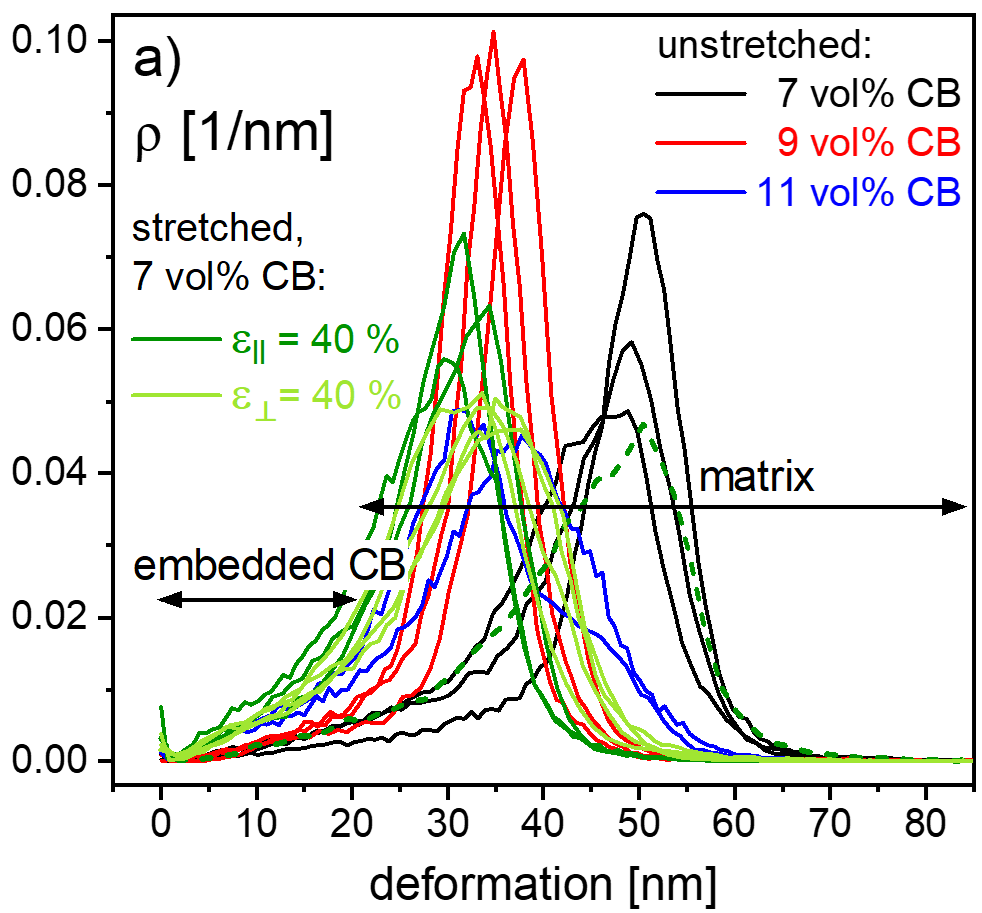}
     \includegraphics[width=0.48\linewidth]{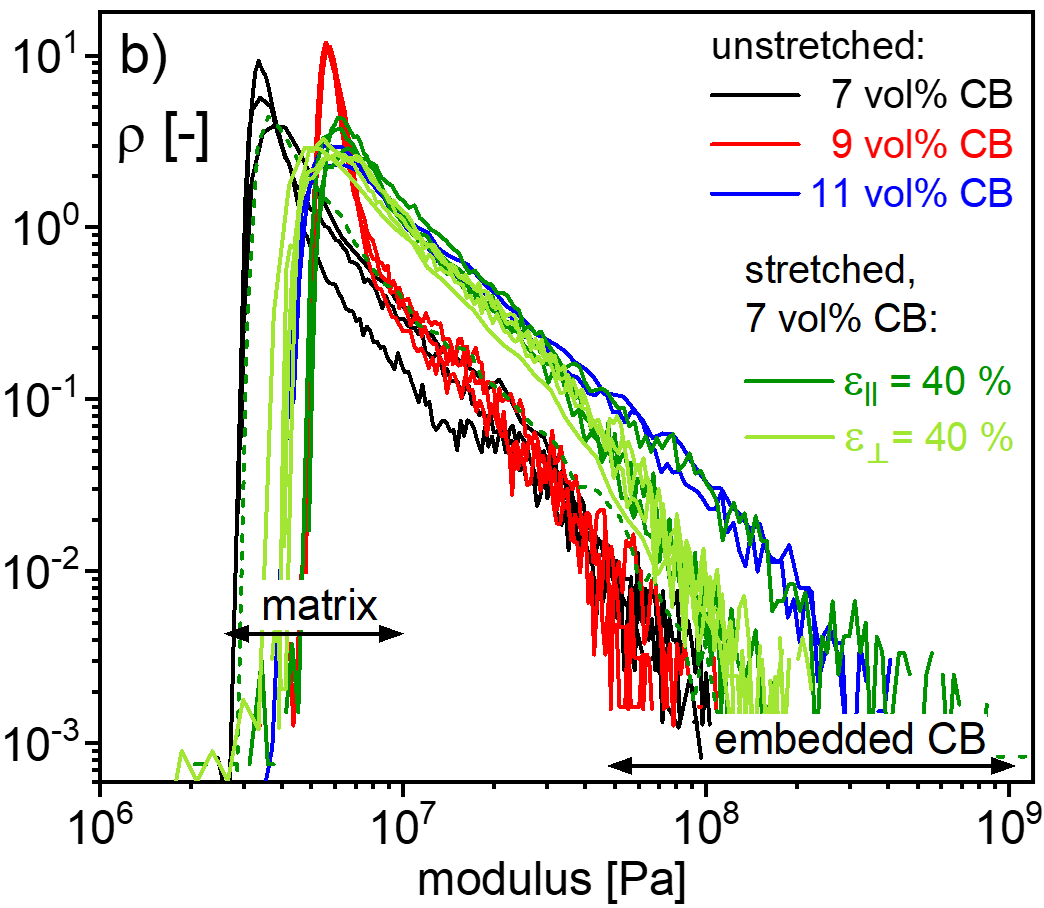}
     \includegraphics[width=0.45\linewidth,trim={0cm 1cm 0cm 0cm}]{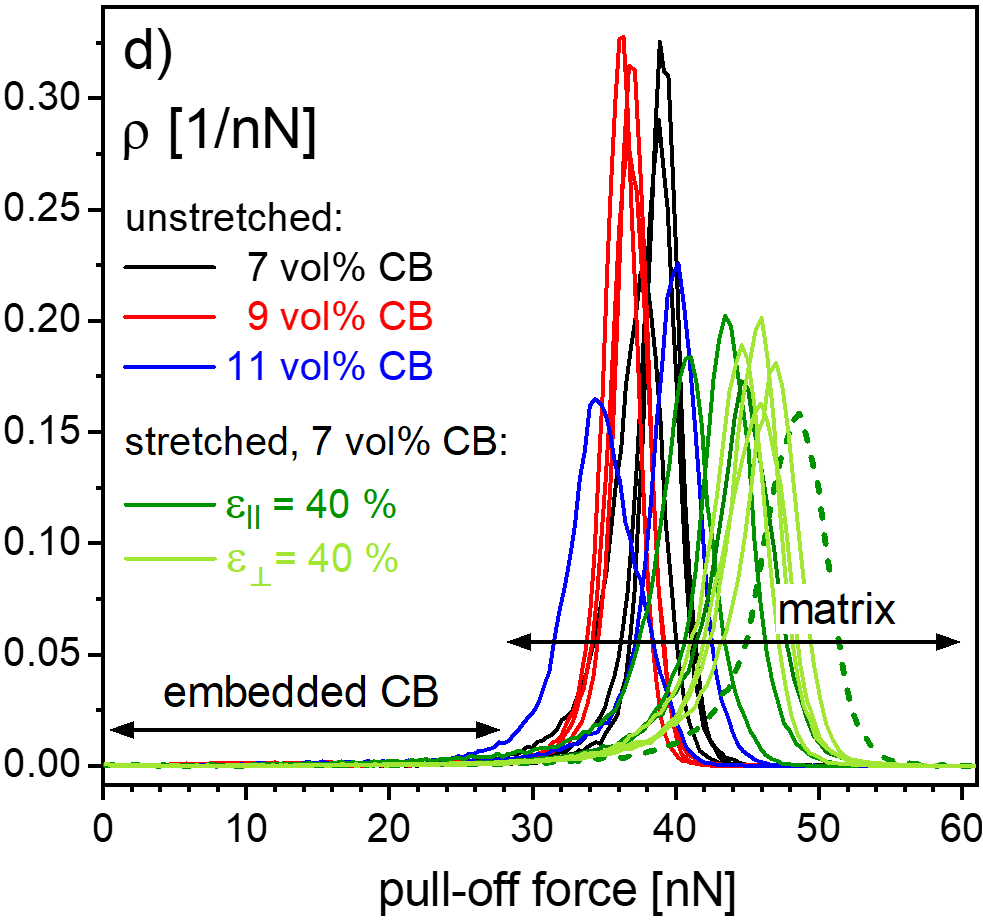}
     \includegraphics[width=0.47\linewidth,trim={0cm 1cm 0cm 0cm}]{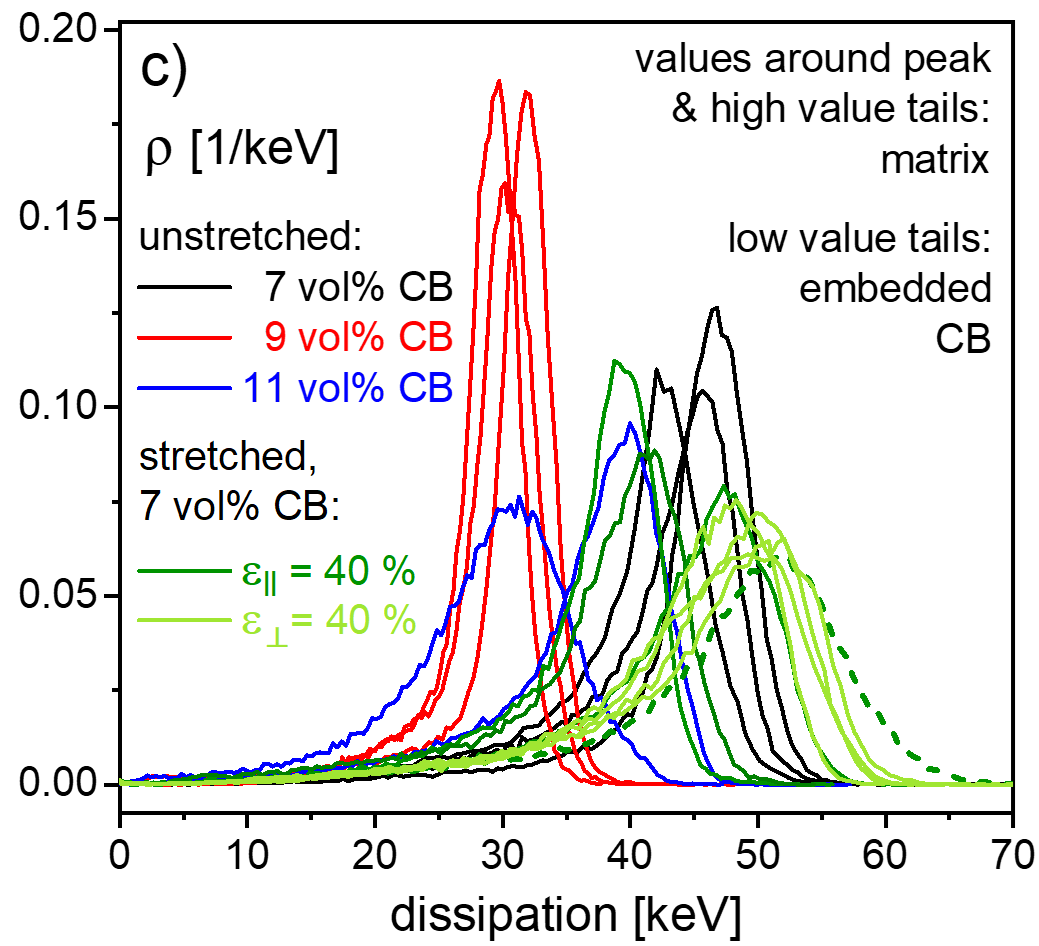}
     \caption{\justifying{Frequency densities of PeakForce QNM signals for unstretched films (7/9/11~vol\% CB) as well as 7 vol\% CB samples stretched 40 \% parallel and perpendicular to the coating direction, respectively (2 – 4 scans of 10x10 µm\textsuperscript{2} per material state). Scans of the unstretched 7 vol\% CB-film and one scan on the 7 vol\% CB-film stretched parallel (green, dashed curves) were acquired with the same SFM probe, the rest of the scans with another probe. The frequency density of the modulus signal is unitless because it is derived from log-scaled values, log(modulus[Pa]), that account for the big differences in stiffness of CB- and matrix-dominated regions.}}
     \label{fig:PF-QNM frequency densities}
 \end{figure}

Frequency densities for unstretched films (7/9/11 vol\% CB) and stretched 7 vol\% CB-samples, along with the rough assignment of CB- and matrix-dominated regions, are shown in Fig. \ref{fig:PF-QNM frequency densities}. Due to the misleading character of the term ‘adhesion’, we designated the values of the adhesion signal by the neutral term ‘pull-off force’ (see Section \ref{General information on PeakForce QNM signals}).

Concerning reproducibility, we first point out that after measuring the unstretched 7 vol\% CB-film and one spot on the 7 vol\% CB sample stretched parallel, the SFM probe had to be replaced (corrupted force-distance curves, probably due to tip wear). For scans with the same SFM probe, different measuring spots of a given composition and material state have very similar (7/9 vol\% CB) or at least similar (11 vol\% CB) frequency densities. In conclusion, measurements are repeatable within a measuring series of the same SFM probe, laser alignment and PeakForce QNM calibrations. For measurements with different SFM probes, variations are bigger, as seen by the curves for the 7 vol\% CB-sample stretched parallel to the coating direction (dashed curve from probe \#1, other curves from probe \#2). This is not surprising as Bruker and other authors \cite{Xu-Recent} state that truly quantitative measurements are difficult to achieve. For example, Bruker reports that the variation in mean modulus values introduced by using different probes and re-calibrating equals up to 25~\% \cite{Pittenger-Quantitative}. Moreover, tip contamination during scanning can vastly alter achieved values, usually increasing the pull-off force and the measured effective modulus \cite{Zimmer-Hygrothermale}. For the data displayed in Fig. \ref{fig:PF-QNM frequency densities}, the variation from using different probes  and re-calibrating is similar to possible differences between the three compositions and between unstretched vs. stretched films. As a result, we cannot discern trends caused by varying CB concentration or by stretching.

Concerning absolute values, we comment on some physically interesting differences between the PeakForce QNM signals for embedded CB particles and the silicone matrix.
As already hinted at, the embedded CB particles are much stiffer than the silicone matrix, with effective modulus values amounting to 50~–~100~MPa and 3~–~10~MPa, respectively. These big differences in stiffness are a direct (nanomechanical) explanation of the macroscopic reinforcing effect of CB evidenced by mechanical characterization. To underline the plausibility of the absolute values, the modulus range for our PDMS-based silicone matches that for other PDMS samples characterized by PeakForce QNM (mean values of some 100~MPa, up to 10~MPa \cite{Pittenger-Quantitative}, compare to values of peaks in Fig. \ref{fig:PF-QNM frequency densities}b: 3~–~6~MPa).

Concerning pull-off force, the adhesion maps (see Fig. \ref{fig:PF-QNM signal maps typical scaling} and discussion in Section \ref{Nanomechanical mapping (PeakForce QNM)_results}) show that the silicon tip adheres more strongly on silicone than it does on carbon. Since CB particles in the adhesion maps appear to be much smaller and more scarce than in the other signals, we can conclude that the pull-off force is dominated by adhesive interactions between the tip and sample surface, and that a large fraction of near-surface CB is coated by a thin layer of silicone.

Concerning dissipation, the much higher values for the silicone matrix can be attributed to its viscoelastic relaxation and higher deformability compared to CB. Both result in a more pronounced pull-off event (more dissipation via dynamic effects and a larger maximal contact area). In addition, the viscoelastic relaxation leads to a larger hysteresis in the repulsive region. Differences in adhesion (silicon tip to silicone vs. to CB) appear to be less relevant since the dissipation values of uncoated portions of CB particles (black spots in the adhesion maps) and coated portions are very similar (compare regions corresponding to the black spots in the adhesion maps with other dark regions in the dissipation maps).

\section{Simulation details} \label{Appendix: Simulation}

\subsection{MPC simulations}

The MPC algorithm alternates between streaming and collision steps. 
It acts on a group of fictitious point particles representing the solvent that are first propagated ballistically in parallel to the standard MD steps. 
In certain time intervals, corresponding to an MPC particle free path of $h=0.1\,a$, MD and MPC particles are sorted into the compartments of a grid.
Momentum is exchanged between solvent and solute in these so-called collision cells during the collision step. 
This is achieved by applying the scheme of Stochastic Rotation Dynamics (SRD), in which all velocities in a cell relative to the center of mass velocity are rotated by an angle of $\pm 90^\circ$ around one of the Cartesian axes. 
Axis and direction of rotation are drawn randomly from a uniform distribution. 
%This step couples solvent and solutes. 
Shifting the grid randomly restores Galilean invariance.
We work in MPC units, i.e., choose $k_\mathrm{B}T = 1$, MPC particle mass $m=1$, MPC collision cell size $a=1$, and derive the time unit $\tau = a\sqrt{m/(k_\mathrm{B}T)}$. 
The CB primary particle mass $M = \rho m$ with MPC particle number density $\rho = 10$ accounts for neutral buoyancy. 
This is a standard choice of parameters, yielding a viscosity of $\eta = 4.6 m/(a \tau)$ and a Schmidt number of $\Sc = 5.3$. 
For our MPC-MD simulations, we employ a modified \cite{chen2018coupling} version of LAMMPS (2 Aug 2023) \cite{LAMMPS, petersen2010mesoscale, in2008accurate, shire2021simulations} and its simple local rescaling thermostat \cite{bolintineanu2012no} that accounts for the shear flow by acting on the velocities relative to the expected flow profile.

For the MD simulation of CB aggregates, we restrain point particles representing the primary particles by harmonic potentials of the form $U(r) = k_b  (r - r_0)^2$ for bonds ($k_b = 450\,k_\mathrm{B}T/\sigma^2$, $r_0=1$), analogously for angles and dihedrals ($k_a = k_d = 100\,k_\mathrm{B}T/\text{rad}^2$). The CB particles are propagated with a velocity Verlet integrator, for which we use a small time step $\Delta t = 10^{-4}\,\tau$ to account for the stiff potentials. Aggregates interact particle-wise with each other, but not with themselves, with a Weeks-Chandler-Anderson potential with $\epsilon = 10\,k_\mathrm{B}T$, $\sigma = 1\,a$. For the nematic order parameters and the orientation distributions, we measure over $1\cdot 10^5\,\tau$ for $\Pe>2$ and $2\cdot 10^5\,\tau$ otherwise.

\subsection{Conductivity computation}

\begin{figure}
\centering
\includegraphics[width=0.5\textwidth]{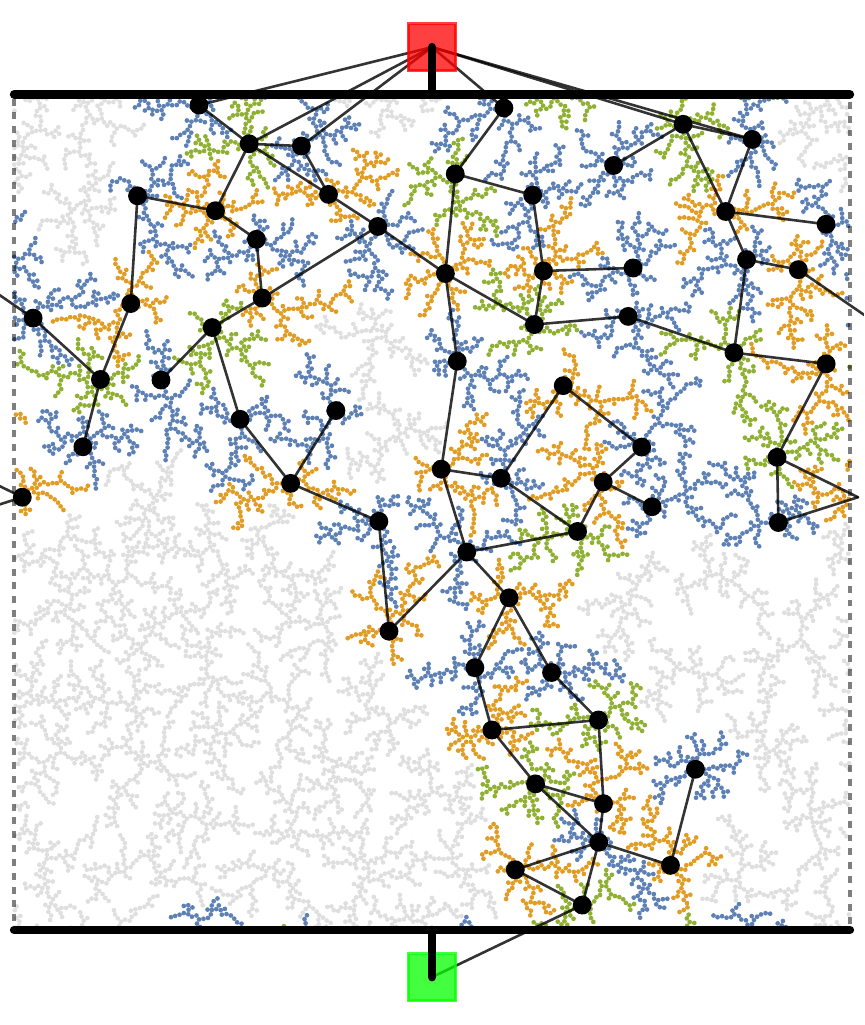}
\caption{\justifying{Illustration of the Kirchhoff network generated from an aggregate system. Conductivity is measured between artificial electrodes (rectangular nodes on top and below the system). Figure taken from \cite{coupette2023percolation}.}}
\label{fig:agg_network}
\end{figure}

The nodes in the Kirchhoff network represent the aggregates, while the edges correspond to connections between aggregates. Fig.~\ref{fig:agg_network} illustrates the construction.
Each connection is weighted by its individual conductance which decays exponentially with the gap between the two primary particles that form the tunneling contact. 

Multiple contacts between the same aggregates are considered in parallel so that they can be absorbed into a single edge weighted by the sum of the individual conductances. 
Tangential aggregates are set to have unit conductance, which is also the maximum conductance between any two aggregates.
A contact at separation $d$, i.e., the threshold distance for contact identification, accounts for $1\%$ of unit conductance to be consistent with discarding longer-range contacts which fixes the conductance model. 
The network is decomposed into bi-connected components relative to the electrodes in order to eliminate dead ends and Wheatstone bridges from the network. 
We choose $d$ as $\frac{\sigma}{2}$ because it induces a percolation threshold just below $10\,\text{vol-}\%$, which is comparable to the values observed experimentally. However, a realistic tunneling distance will be much smaller just like the aggregates are realistically significantly bigger, amounting to a similar critical volume fraction. However, our observations for the anisotropic conductivity ratio are robust against larger variations of $d$ with only the absolute values of conductivities changing by orders of magnitude.
\end{appendices}
\end{spacing}

\end{document}